\documentclass[12pt]{iopart}
\usepackage[utf8x]{inputenc}
\usepackage[UKenglish]{babel}
\expandafter\let\csname equation*\endcsname\relax
\expandafter\let\csname endequation*\endcsname\relax
\usepackage{amsmath}
\usepackage{iopams}
\usepackage{amssymb}
\usepackage{fullpage}
\usepackage{bm}
\usepackage{bbold}
\usepackage{appendix}
\usepackage{graphicx}
\usepackage{color}
\usepackage{tikz}
\usetikzlibrary{calc,decorations.markings}
\newcommand{\be}{\begin{equation}}
\newcommand{\ee}{\end{equation}}

\newcommand{\bea}{\begin{eqnarray}}
\newcommand{\eea}{\end{eqnarray}}

\newcommand{\muh}{\hat{\mu}}
\newcommand{\dx}{\delta x}

\newcommand{\dR}{\delta R}

\newcommand{\xe}{\Xi^{\rm eff}}
\newcommand{\lhe}{l_i^{\rm eff}(t)}
\newcommand{\he}{\bm{h}^{\rm eff}}
\newcommand{\psie}{\psi_i^{\rm eff}(t)}
\newcommand{\che}{\hat{C}_i^{\rm eff}(t,t')}
\newcommand{\rhe}{\hat{R}_i^{\rm eff}(t,t')}
\newcommand{\bhe}{\hat{B}_i^{\rm eff}(t,t')}
\newcommand{\Ga}{G_\alpha}
\newcommand{\Gha}{\hat{G}_\alpha}
\newcommand{\Gexp}[1]{G^{#1}}
\newcommand{\Ha}[1][\alpha]{\mathcal{H}_#1}
\newcommand{\Hint}{\mathcal{H}_{\rm int}}
\newcommand{\Xa}{\Xi_\alpha}
\newcommand{\mv}{\bm{m}}
\newcommand{\hv}{\bm{h}}
\newcommand{\hexp}[1]{\bm{h}^{#1}}
\newcommand{\hhexp}[1]{h^{#1}}
\newcommand{\Sv}{\bm{S}}
\newcommand{\eff}{^{\text{eff}}}
\usepackage{cite}
\numberwithin{equation}{section}
\allowdisplaybreaks[1]

\begin{document}
\title[Extended Plefka Expansion]{Extended Plefka Expansion for Stochastic Dynamics}
\author{B Bravi$^1$, P Sollich$^1$ and M Opper$^2$}
\address{$^1$ Department of Mathematics, King's College London, Strand, London, WC2R 2LS UK}
\address{$2$ Department of Artificial Intelligence, Technische Universit\"{a}t Berlin, Marchstra{\ss}e 23,
Berlin 10587, Germany}
\ead{barbara.bravi@kcl.ac.uk, peter.sollich@kcl.ac.uk and manfred.opper@tu-berlin.de}
\date{}

\begin{abstract}
We propose an extension of the Plefka expansion, which is well known for the dynamics of discrete spins, to stochastic differential equations with continuous degrees of freedom and exhibiting generic nonlinearities. The scenario is sufficiently general to allow application to e.g.\ biochemical networks involved in metabolism and regulation. The main feature of our approach is to constrain in the Plefka expansion not just first moments akin to magnetizations, but also second moments, specifically two-time correlations and responses for each degree of freedom. The end result is an effective equation of motion for each single degree of freedom, where couplings to other variables appear as a self-coupling to the past (i.e. memory term) and a coloured noise. This constitutes a new mean field approximation that should become exact in the thermodynamic limit of a large network, for suitably long-ranged couplings. For the analytically tractable case of linear dynamics we establish this exactness explicitly by appeal 
to spectral methods of Random Matrix Theory, for Gaussian couplings with arbitrary degree of symmetry.
\end{abstract}
\submitto{\jpa}
\maketitle

\section{Introduction}
Stochastic Differential Equations (SDEs) with continuous variables are a well-established tool to describe the dynamical behaviour of a variety of systems, in areas ranging from physics and chemistry to biology and engineering \cite{vankampen,gardiner}: they are used frequently, for example, for dynamical modelling of intracellular kinetics and biochemical networks \cite{systems}.

In the context of network studies, in particular with regard to applications in systems biology, a major task is model simplification \cite{okino,apri}, using model reduction strategies that should retain as much as possible of the qualitative dynamical information. In addition one requires techniques for inferring model parameters from observed data,  
since experimental uncertainties on parameters, resulting e.g.\ from the fact that some dynamical variables may not be observed, can crucially affect the predictions of dynamical models \cite{achcar}.

The application of approaches based on statistical mechanics and spin glass theory has a long history \cite{virasoro}. In particular,
mean field (MF) methods have emerged as powerful tools for characterizing statistical quantities in systems where the combinatorial complexity of exact calculations rules out a tractable description \cite{meanfield}. From the theoretical point of view, further motivation for the use of mean field methods comes from the fact that they can often be proved to retrieve the exact solution in an appropriate limit, typically involving high network connectivity and/or weak couplings.

The so called  ``Plefka expansion" for the Sherrington-Kirkpatrick (SK) \cite{SK} model was introduced by Plefka \cite{plefka} as a convenient method to derive MF  
equations and their more refined analogue, the TAP equations \cite{TAP}. The advantage of the method, essentially an expansion of the Gibbs free energy in powers of the interaction 
strength, is that it does not rely on an average over interactions drawn from some statistical ensemble. This makes it potentially useful in applications to e.g.\ biology, where it is 
generally a {\em specific} network that is of interest.

Roudi and Hertz \cite{roudi} applied the Plefka expansion to the problem of approximating spin-glass dynamics: in this case, variables are not single spins but entire time histories of each spin.
They developed a dynamical theory that relates mean magnetizations, potentially time varying fields and quenched couplings for two versions of SK model kinetics (synchronous and asynchronous updates, respectively). Using the generating functional approach, the (naive) MF and TAP dynamical equations were retrieved as first and second orders of a power expansion in analogy with the equilibrium Plefka expansion for the Gibbs free energy. In more detail, the logarithm of the generating functional for the dynamics
plays the role of the equilibrium free energy: performing the Legendre transform w.r.t.\ the real and auxiliary fields one obtains the dynamical equivalent of the Gibbs free energy and then can expand for weak couplings. Importantly, as long as the generating functional is by definition dependent only
on fields that act linearly on the degrees of freedom, this expansion will closely resemble the standard Plefka approach and only the first moments of the resulting probability measure over trajectories will be fixed.

The aim of our paper is two-fold. First we want to introduce an improvement, taylored to continuous degrees of freedom, of the approximation strategy outlined above; we call the improved method an ``extended" Plefka expansion.
The dynamical model is a set of stochastic differential equations for continuous degrees of freedom and with generic nonlinear couplings between them. The basic idea of the extension 
that we propose is to include among the set of order parameters all second moments, i.e.\ two-time correlations and responses, for each degree of freedom. Expanding up to second order 
in interaction strength then provides a mean field description where couplings between trajectories are replaced by a coupling to the past (i.e.\ a memory term) and 
a coloured noise. 

Our second aim is an analytical investigation of a solvable limit, which concerns large networks with linear dynamics. This partly serves the purpose of verifying explicitly a case 
where the approximation becomes exact, but the calculation also provides additional insight how the dynamical behaviour of correlations and responses depends on the symmetry of the couplings. 
We show that the exact thermodynamic limit is recovered from the approximate
equations for any degree of symmetry, i.e.\ irrespective of whether the system reaches an equilibrium stationary state. This keeps the analysis as general as possible and suggests multiple possible applications, for  
example in neural networks and gene expression where couplings are typically asymmetric.

The paper is organized as follows: after recalling the expansion conceived by Plefka in section \ref{sec:PE}, we introduce in section \ref{sec:EPE} the basic functional integral approach that provides 
the framework within which we build the extended Plefka expansion for dynamics. In sections \ref{sec:MF} and \ref{sec:TAP} we present and discuss the derivation of the approximate dynamical 
equations from the functional integral. In sections \ref{sec:LD}, \ref{sec:LD1} and \ref{sec:TLLD} we apply the approximation to the particular case of a linear dynamics, which is analytically tractable both in the static and dynamic scenario. In section \ref{sec:ED} we resort to Random Matrix Theory and related spectral methods \cite{mehta} to average the exact dynamics over the disordered interactions, in the limit of an infinitely large sample and in the stationary regime.
This allows us to derive expressions for correlations and responses in Laplace space, and comparison with the predictions of the extended Plefka approximation shows perfect agreement. This 
confirms and strenghtens the theoretical justification of our method. In section \ref{sec:PS} we study in more detail the qualitative features of the dynamics, in particular non-exponential relaxation 
behaviour that manifests as low-frequency power law tails in the power spectra. Finally, an explicit analytical characterization of correlations and responses in the temporal domain can be found in the limit of symmetric and antisymmetric couplings and is discussed briefly in section \ref{sec:TD}.

\section{Plefka Expansion}
\label{sec:PE}
We briefly summarize the main steps of the ``Plefka expansion'' introduced by Plefka \cite{plefka}, using, as in the
original paper, the Sherrington-Kirkpatrick (SK) \cite{SK} model as an example.
The SK model of a spin glass consists of $N$ Ising spins ($S_i=\pm 1$) with Hamiltonian
\begin{equation}
 \mathcal{H}= \frac{1}{2}\sum_{i\neq j}J_{ij}S_iS_j + \sum_i h_i^{\rm ext} S_i
\label{eq:Hdef}
\end{equation}
In the SK model, specifically, the interactions are symmetric (i.e.\ $J_{ij}=J_{ji}$) and infinitely long-ranged, with the $J_{ij}$
for $i<j$ chosen as independent Gaussian variables of mean zero and
variance $1/N$, though these properties are not required to write down
the general expansion. Note that the left hand side of (\ref{eq:Hdef})
would conventionally be written as $-\beta \mathcal{H}$ with $\beta$ the inverse temperature, but we omit factors of $-\beta$ here and below as we do not need them in the application to dynamics.
In order to construct the Plefka expansion one introduces a parameter $\alpha$ controlling the interaction strength, defining a modified Hamiltonian as
\begin{equation}
\Ha = \frac{\alpha}{2}\sum_{i\neq j}J_{ij}S_iS_j + \sum_i h^{\rm ext}_i S_i
 \end{equation}
The full interacting Hamiltonian is then $\Ha[1]=\mathcal{H}$, while $\Ha[0]$ is the Hamiltonian of a non-interacting system. 
The Gibbs free energy $\Ga$ is now defined as the free energy
subject to a constraint on certain averages, typically the magnetizations
$m_i=\langle S_i\rangle$
\begin{equation}
\Ga(\mv) = \text{extr}_{\hv} \Gha(\mv, \hv)
\end{equation}
with
\begin{equation}
\Gha(\mv, \hv)= \ln \text{Tr}\,\text{e}^{\Xa} 
\end{equation}
and
\begin{equation}
\label{eq:s_m_zero}
\Xa = \Ha + \sum_i h_i(S_i - m_i)
\end{equation}
One can write 
\begin{equation}
\Ga(\mv) = \text{extr}_{\hv} \left(
\ln \text{Tr}\,\text{e}^{\Ha + \sum_i h_i S_i} 
- \sum_i h_i m_i\right)
\end{equation}
and this shows that $\Ga$ is the Legendre transform of a Helmholtz free energy -- the first term in the brackets -- that depends on
the auxiliary fields $h_i$. The extremization condition over the $h_i$
gives
\begin{equation}
m_i = \langle S_i \rangle
\label{eq:m_equals_average_S}
\end{equation}
and this ensures that the $m_i$ have the intended meaning. The average here is over the distribution of states $P(\Sv) \propto \text{e}^{\Xa}$. This is biased away from the Boltzmann distribution $(1/Z)\text{e}^{\Ha}$ by the factor 
$\text{e}^{\hv\cdot\Sv}$ involving the auxiliary fields $h_i$. We will denote the fields that produce the desired values of the magnetizations $\mv$ by $\hv_\alpha(\mv)$, where the subscript emphasizes the dependence on the interaction strength $\alpha$.
The fields $\hv_\alpha$ can be deduced as derivatives of $\Ga$, once this is known. Explicitly, because of
the condition (\ref{eq:m_equals_average_S}), the variation of the fields $\hv_\alpha$ with $\mv$ does
not contribute to the $\mv$-derivative of $\Ga$, so that
\begin{equation}
\frac{\partial \Ga}{\partial m_i} = -h_{i\alpha}
\end{equation}
as expected on general grounds from the Legendre transform
definition of $\Ga$. The Gibbs free energy becomes equal to the
unconstrained equilibrium free energy when the fields $h_{i\alpha}$ vanish, so
that the condition for the equilibrium magnetizations is simply
\begin{equation}
\frac{\partial \Ga}{\partial m_i} = 0
\end{equation}
The formalism so far is generic. In the Plefka expansion, the
interacting part of the Hamiltonian is treated perturbatively by
expanding the Gibbs free energy in powers of $\alpha$, typically to
first or second order
\begin{equation}
\Ga = \Gexp{0}+\alpha\Gexp{1}+\frac{\alpha^2}{2}\Gexp{2}+\ldots
\end{equation}
where $\Gexp{k} = (\partial / \partial \alpha)^k \left. \Ga \right|_{\alpha=0}$.
The fields $h_{i\alpha}= -\partial \Ga / \partial m_i $ can be expanded analogously 
\begin{equation}
\hv_{\alpha} = \hexp{0} + \alpha \hexp{1} +\frac{\alpha^2}{2} \hexp{2} + \ldots
\end{equation}
To the second order, the equilibrium condition $\hv_\alpha=0$ for the order
parameters $\mv$ is then given by
\be 
0 = \hexp{0} + \alpha \hexp{1} +\frac{\alpha^2}{2} \hexp{2}
\label{eq:equilibrium_nonzero_alpha}
\ee 
In applications to equilibrium spin systems, the non-interacting Gibbs
free energy $\Gexp{0}$ can often be found explicitly, e.g.\ for our
Ising spin example
\begin{equation}
G^0= -\sum_i\left[\frac{1+m_i}{2}\ln{\bigg(\frac{1+m_i}{2}\bigg)} +\frac{1-m_i}{2}\ln{\bigg(\frac{1-m_i}{2}\bigg)}\right] + \sum_i h^{\rm ext}_i m_i
\end{equation}
In dynamical problems, finding $\Gexp{0}$ explicitly
is often awkward but can be avoided by noting that 
in order to obtain a certain value of $\mv$ at $\alpha=0$ requires
a field $\he=\hexp{0}$. The equilibrium condition
(\ref{eq:equilibrium_nonzero_alpha}) for nonzero $\alpha$ can then be
rewritten as
\be 
\label{eq:zerofields}
\he = -\alpha \hexp{1} -\frac{ \alpha^2}{2} \hexp{2}
\ee 
This expression gives us the effective fields $\he$ that produce the
{\em same} magnetizations $\mv$ in the non-interacting system as
at equilibrium in the interacting system. To obtain the equilibrium
condition for the interacting system, one then only needs to combine
this with the relation between magnetization and field in the
non-interacting system, which for Ising spins reads simply
\be 
m_i = \tanh(h^{\rm ext}_i + h_i\eff)
\ee 
To carry out the actual calculation of the first and second order
Plefka free energies $\Gexp{1}$ and $\Gexp{2}$, one notes first that
$\Ga(\mv)=\Gha(\mv,\hv_\alpha(\mv))$, hence
\be
\frac{\partial \Ga}{\partial \alpha}
= \frac{d \Gha}{d \alpha} = \left\langle \frac{d \Xa}{d\alpha}\right\rangle_{\alpha}
\label{eq:G1_first_version}
\ee
where we use $(d/d\alpha)$ to indicate a total derivative that
includes the $\alpha$-dependence of $\hv_\alpha$. On the other hand
\eqref{eq:s_m_zero} shows that in $\Xa$ each field
$h_{i\alpha}$ multiplies $S_i-m_i$, whose average vanishes,
so this $\alpha$ dependence drops out and one has simply
\be
\frac{\partial \Ga}{\partial \alpha}
= \left\langle \Hint \right\rangle_{\alpha}
\label{eq:G1_second_version}
\ee
where $\Hint = \partial \Ha/\partial\alpha$ is the interacting part of the original Hamiltonian.
Evaluating the average in the non-interacting system ($\alpha=0$) then
gives $\Gexp{1}=\langle \Hint \rangle_0$, and by derivation
$\hexp{1}$. For the SK model, one finds in this way $\Gexp{1}=
(1/2)\sum_{i\neq j}J_{ij}m_im_j$ and $h^{1}_i = - \partial
\Gexp{1}/\partial m_i = - \sum_{j\neq i} J_{ij}m_j$. To first order
the effective field is then $h\eff_i = -\alpha \hhexp{1}_i = \alpha
\sum_{j\neq i} J_{ij}m_j$ and the equilibrium condition $m_i =
\tanh(h^{\text{ext}}_i + \alpha \sum_{j\neq i}J_{ij}m_j)$ has the
familiar mean-field form.
For the second order one has in general
\begin{eqnarray}
\frac{\partial^2 \Ga}{\partial \alpha^2}&= \frac{d^2 \Gha}{d \alpha^2}=\notag \\
&= \left\langle \frac{d^2 \Xa}{d\alpha^2}\right\rangle_{\alpha}
+ \biggl\langle \left(\frac{d \Xa}{d\alpha}\right)^2
\biggr\rangle_{\alpha} 
- \biggl\langle \frac{d \Xa}{d\alpha}\biggr\rangle_{\alpha}^2
\end{eqnarray}
The first term vanishes because $\partial^2 \Ha/\partial\alpha^2 = 0$
and because $\partial^2 h_{i\alpha}/\partial \alpha^2$ is multiplied
again by a vanishing average. Evaluating at $\alpha=0$ then gives (as discussed in \cite{roudi})
\bea
\label{eq:roudi}
\Gexp{2}
&=& \biggl\langle \left(\delta \frac{d \Xa}{d\alpha}\right)^2\biggr\rangle_0
\eea
where
\be
\delta \frac{d \Xa}{d\alpha}
= \frac{d \Xa}{d\alpha} - \left\langle \frac{d \Xa}{d\alpha}\right\rangle_0
= \Hint - \langle \Hint\rangle_0 + \hexp{1}\cdot(\Sv-\mv)
\ee
From $\Gexp{2}$ one finds $\hexp{2}$ by taking $\mv$-derivatives
again, and in principle this process can be iterated to higher order. The first order gives a MF approximation as shown above, 
while at second order one retrieves what are known as the TAP equations for the SK-model \cite{SK}.

\section{Extended Plefka expansion}
\label{sec:EPE}
We start from the dynamical equations
\begin{equation}
\frac{dx_i(t)}{dt} = - \lambda_{i}x_i(t)+\phi_i(\bm{x}(t))  +\xi_i(t)
\end{equation}
for a set of $N$ continuous (real-valued) degrees of freedom $x_i$ ($i=1,\ldots,N$) evolving in time $t$. The $x_i$ may represent e.g.\ concentrations of chemical species in a biochemical reaction network, or 
deviations of such concentrations from steady state values. On the r.h.s., $\phi_i(\bm{x}(t))$ is a generic function of the vector $\bm{x}(t)=\lbrace x_i(t) \rbrace$ of all concentrations and determines the drift of $x_i$.
In the biochemical context it gives the rate of change in $x_i$ due to reactions with other species and includes the relevant reaction rates. A term $-\lambda_i x_i$ has been included that drives each $x_i$ back to zero, with $\lambda_i$ having the meaning of a decay rate. Finally, $\xi_i(t)$ is  Gaussian white noise with the properties
\begin{equation}
\langle\xi_i\rangle=0 \qquad \langle \xi_i(t) \xi_j(t')\rangle= \Sigma_{ii} \delta_{ij}\delta(t-t')
\end{equation}
The Kronecker delta $\delta_{ij}$ signifies that each variable $x_i$ has independent noise acting on it. Correlations in the noise could be allowed for by extending the matrix $\Sigma_{ii}\delta_{ij}$ to one having nonzero off-diagonal entries, but become difficult to express in terms of the local parameters that define the core of the extended Plefka expansion, as will be explained below.

After discretizing time in elementary time steps $\Delta$, a dynamical partition function for this system can be written in the Martin--Siggia--Rose--Janssen--De Dominicis (MSRJD) 
functional integral formalism \cite{martin}, \cite{janssen}, \cite{dedominicis} 
\begin{eqnarray}
Z&=&\bigg\langle \int\prod_{it}dx_i(t)\delta\big( x_i(t+\Delta)-x_i(t)-\Delta[-\lambda_i x_i(t)+\phi_i(\bm{x}(t))+\xi_i(t)] \big)\bigg\rangle_{\bm{\xi}} =\notag \\
&=&\bigg\langle\int \prod_{it}\frac{dx_i(t)d\hat{x}_i(t)}{2\pi}\text{e}^{\text{i}\hat{x}_i(t)\left(x_i(t+\Delta)-x_i(t)-\Delta[-\lambda_i x_i(t)+\phi_i(\bm{x}(t)) +\xi_i(t)]   \right)} \bigg\rangle_{\bm{\xi}}
\end{eqnarray}
We use the It\^o convention \cite{vankampen} to discretize the noise, where $\xi_i(t)$ above is to be read as the average of the continuous-time noise over the time interval $[t,t+\Delta]$, which has covariance \begin{equation}
\langle {\xi}_i(t){\xi}_i(t') \rangle=\frac{1}{\Delta}\Sigma_{ii}\delta_{tt'}
\end{equation}
Here $\delta_{tt'}/\Delta$ is the discrete-time replacement of $\delta(t-t')$. The average over the white noise can then be performed by applying a standard Gaussian identity
\begin{equation}
\langle \text{e}^{\text{i}\Delta \hat{\bm{x}}^{\rm T} \cdot {\bm{\xi}}} \rangle_{\bm{\xi}}= \text{e}^{-\Delta\,\bm{\hat{x}}^{T}\bm{\Sigma}\bm{\hat{x}}/2}
\end{equation}
To develop a Plefka expansion, we now need to consider which averages should be constrained in the relevant Legendre transform. By reinterpreting the static TAP equations from the perspective of 
a cavity argument \cite{romanobattistin}, one would obtain marginals where the covariance of the cavity field and a quadratic term for the spins is present. These are effectively constant in the 
case of Ising spins ($s_i^2=1$) but should be explicitly taken into account for continuous variables (even in a static problem) and for tracking time dependencies (see \cite{romanobattistin} for 
spin dynamics).

Let us now introduce some shorthands to explain in intuitive terms the logic beyond the ``extended'' Plefka expansion, connecting it to the version for equilibrium systems outlined in section \ref{sec:PE}. We denote
\begin{subequations}
\label{shorthands}
\begin{align}
\hat{\bm{m}}&=\lbrace \bm{x}, -\text{i}\hat{\bm{x}}, \bm{x}\bm{x}, -\text{i}\hat{\bm{x}}\bm{x}, \text{i}\hat{\bm{x}}\text{i}\hat{\bm{x}}  \rbrace\\
\bm{m}&=\lbrace \bm{\mu},-\text{i}\hat{\bm{\mu}},\bm{C}, \bm{R}, \bm{B}\rbrace\\
\bm{h}_{\alpha}&=\lbrace \bm{\Psi}_{\alpha}, \bm{l}_{\alpha}, \hat{\bm{C}}_{\alpha}, \hat{\bm{R}}_{\alpha}, \hat{\bm{B}}_{\alpha} \rbrace
\end{align}
\end{subequations}
Here $\hat{\bm{m}}$ is a compact notation for the quantities whose averages we will constrain, consisting of the $x_i(t)$, $\text{i}\hat{x}_i(t)$ and all their products involving the same degree of freedom or ``site'' $i$. It is the inclusion of these products that extends our approach beyond the standard applications of the Plefka method, where only first order moments such as magnetizations are constrained. We indicate by $\bm{m}$ the constrained values of the relevant averages, which are the order parameters of the theory, and by $\bm{h}_\alpha$ the conjugate fields. $\bm{\mu},\hat{\bm{\mu}},\bm{C}, \bm{R}, \bm{B}$ summarize the various groups of order 
parameters defined as follows
\begin{subequations}
\label{eq:moments}
\begin{align}
\mu_i(t)&= \langle x_i(t) \rangle_{\alpha}\\
\hat{\mu}_i(t)&= \langle \hat{x}_i(t) \rangle_{\alpha} \\
C_{i}(t, t')&=  \langle x_i(t)x_i(t')\rangle_{\alpha} \\
R_{i}(t', t)&=  -\text{i}\langle\hat{x}_i(t)x_i(t') \rangle_{\alpha} \\
B_{i}(t, t')&= - \langle\hat{x}_i(t)\hat{x}_i(t')\rangle_{\alpha}
\end{align}
\end{subequations}
We denote the corresponding groups of conjugate fields by $\bm{\Psi}_{\alpha}, \bm{l}_{\alpha}, \hat{\bm{C}}_{\alpha}, \hat{\bm{R}}_{\alpha}, \hat{\bm{B}}_{\alpha}$.

The second order quantities we are constraining involve firstly the (disconnected, local) two-time correlation functions $C_i(t,t')$. From general results for MSRJD path integrals \cite{coolen} it follows that $R_{i}(t', t)$ has the meaning of a local response of $x_i(t')$ to a perturbing field $-\text{i}\hat{x}_i(t)$ applied at some earlier time; it should therefore be non-vanishing only for $t'>t$. $B_i(t,t')$, finally, is expected to vanish for all times $t$ and $t'$, as is $\hat\mu_i(t)$; both follow from the fact that the dynamical partition function remains equal to unity when generating terms linear in $\hat x_i(t)$ are added in the exponent (we refer to \cite{coolen} for a derivation from the normalization condition).

To define the Plefka free energy, note that after the noise average has been carried out, our partition function can be written in the form $Z=\int D\bm{x} D\hat{\bm{x}}\, \text{e}^{\mathcal{H}_{\alpha}}$ with a suitable Hamiltonian (or action) $\mathcal{H}_\alpha$ for the stochastic dynamics. Here $D\bm{x} D\hat{\bm{x}}$ is a shorthand for  the integral $\prod_{it}\frac{dx_i(t)d\hat{x}_i(t)}{2\pi}$ and corresponds to the trace over spins. As in the equilibrium calculation one now defines the Plefka energy $G_{\alpha}$ as
\begin{equation}
\label{eq:gammaxi}
G_{\alpha}(\bm{m}) =\hat{G}_{\alpha}(\bm{m}, \bm{h}_{\alpha}(\bm{m})) =\ln{ \int D\bm{x} D\hat{\bm{x}}\, \text{e}^{\Xi_{\alpha}}}
\end{equation}
where
\begin{equation}
 \Xi_{\alpha}=\mathcal{H}_{\alpha}+ \bm{h}_{\alpha}\cdot (\hat{\bm{m}}-\bm{m})
\end{equation}
Explicitly, one has for our system and with the extended set of Plefka order parameters
\begin{eqnarray}
\fl\Xi_{\alpha}=\sum_{it} \text{i}\hat{x}_i(t)\big(x_i(t+\Delta)-x_i(t)+\Delta\lambda_i x_i(t)-\alpha\Delta\phi_i(\bm{x}(t))\big)+ \Delta\sum_{it}\psi_{i\alpha}(t)\big(x_i(t)-\mu_i(t)\big)+\notag\\
\fl-\Delta\sum_{it}l_{i\alpha}(t)\big(\text{i}\hat{x}_i(t)-\text{i}\hat{\mu}_i(t)\big)+\Delta^2\sum_{itt'}\hat{C}_{i\alpha}(t,t')\big(x_i(t)x_i(t')-C_i(t,t')\big)+\notag\\
\fl+\Delta^2\sum_{itt'}\hat{R}_{i\alpha}(t,t')\big(-\text{i}\hat{x}_i(t)x_i(t')-R_i(t',t)\big)+\frac{\Delta^2}{2}\sum_{itt'}\hat{B}_{i\alpha}(t,t')\big(-\hat{x}_i(t)\hat{x}_i(t')-  B_{i}(t,t')\big)+\notag\\
\fl-\frac{\Delta}{2}\sum_{it}\Sigma_{ii}\hat{x}_i(t)\hat{x}_i(t)
\end{eqnarray}
where the first and last terms constitute the Hamiltonian $\mathcal{H}_\alpha$. Note that we have inserted powers of $\Delta$ in such a way as to keep the fields of order unity in the continuous time limit $\Delta\to 0$.
The parameter $\alpha$ characterizes the strength of the interactions as in the equilibrium case, here via $\phi_i$;  the linear self-interaction via $-\lambda_i x_i$ is tractable and so is left as part of the non-interacting baseline. Our aim will be to use a second-order Plefka expansion to derive an effective non-interacting description of our system, where the interactions between variables are replaced by additional coloured noise and a coupling of each variable to its past.

In analogy with the equilibrium expansion, the fields $\bm{h}_{\alpha}$ are determined by extremization of $\hat{G}_\alpha$. Once $G_\alpha$ has been found, the fields can be retrieved from $\bm{h}_{\alpha}=-\partial G_{\alpha}/\partial \bm{m}$ and order parameters of the original system dynamics can be found from the condition $\bm{h}_{\alpha}=0$. Split into the various order parameter groups, the derivatives of $G_{\alpha}$ read
\begin{subequations}
\label{fields}
\begin{align}
\psi_{i\alpha}(t)&=-\frac{1}{\Delta}\frac{\partial G_{\alpha}}{\partial \mu_i(t)}\\
-\text{i}l_{i\alpha}(t)&=-\frac{1}{\Delta}\frac{\partial G_{\alpha}}{\partial (\hat{\mu}_i(t))}\\
\hat{R}_{i\alpha}(t,t') &= -\frac{1}{\Delta^2}\frac{\partial G_{\alpha}}{\partial R_i(t',t)}\\
\hat{C}_{i\alpha}(t,t') &= -\frac{1}{\Delta^2}\frac{\partial G_{\alpha}}{\partial C_i(t,t')}\\
\hat{B}_{i\alpha}(t,t') &= -\frac{1}{\Delta^2}\frac{\partial G_{\alpha}}{\partial B_{i}(t,t')}
\end{align}
\end{subequations}
We now proceed with the Plefka expansion of $G_{\alpha}$ around $\alpha=0$ up to second order, and define a set of effective fields $\he$ as in \eqref{eq:zerofields}. These provide the effective non-interacting 
description of the true interacting dynamics, whereby with these fields at $\alpha=0$ the order parameters have the same values as in the interacting system. As the $\he$ themselves depend on the order parameters, 
this typically leads to nonlinear self-consistency equations, which are the analogues of the MF and TAP equations for the SK model. 

The above makes clear why we have introduced only fields depending on a single site: this assumption guarantees that the effective dynamics will be non-interacting. We also see now why correlations between the noises $\xi_i$ affecting the different $x_i$ would complicate matters: the correlations $C_{ij}(t,t')$ would be non-local even at $\alpha=0$, and determined only in a very indirect way from the local order parameters $C_i(t,t')$.
In the application to biochemical reaction networks there generally are non-trivial noise correlations as discussed in section \ref{sec:conclusions} below, and further work would be required to understand how best to deal with those.

\subsection{Structure of the non-interacting problem}
In the logic explained above, the intractable part of the interactions becomes condensed into local fields that describe the effective single-site dynamics. These effective fields
$\psie, \lhe, \rhe$, $\che, \bhe$ appear in the corresponding effective action $\xe$
\begin{eqnarray}
\fl\xe=\sum_{it} \text{i}\hat{x}_i(t)\bigg(x_i(t+\Delta)-x_i(t)+\Delta\lambda_i x_i(t)\bigg)-\frac{\Delta}{2}\sum_{it}\Sigma_{ii}\hat{x}_i(t)\hat{x}_i(t)
+\Delta\sum_{it}\psie\bigg(x_i(t)-\mu_i(t)\bigg)+\notag \\
\fl-\Delta\sum_{it}\lhe\bigg(\text{i}\hat{x}_i(t)-\text{i}\hat{\mu}_i(t)\bigg)+\Delta^2\sum_{itt'}\che\bigg(x_i(t)x_i(t')-C_i(t,t')\bigg)+\notag \\
\fl+\Delta^2\sum_{itt'}\rhe\bigg(-\text{i}\hat{x}_i(t)x_i(t')-R_i(t',t)\bigg)+\frac{\Delta^2}{2}\sum_{itt'}\bhe\bigg(-\hat{x}_i(t)\hat{x}_i(t')-  B_{i}(t,t')\bigg)
\end{eqnarray}
To get the generic self-consistency equations for our order parameters, we should in principle
evaluate the
averages $\mu_i(t)$, $\muh_i(t)$, $C_i(t,t')$, $R_i(t',t)$ and $B_i(t,t')$ for this action. The result is the analogue of what for an equilibrium spin problem is $m_i=\tanh(h^{\rm ext}_i + h_i\eff)$.

To simplify this procedure, one can make the natural (see above) assumptions that the solution of the self-consistency equations will obey $\hat{\mu}_i(t)= 0$, $B_i(t,t')= 0$ and $R_i(t,t')=0$ for $t'\geq t$;
the vanishing of the response at equal times is a generic consequence of the It\^o discretization. We will have to check that these assumptions are self-consistent. As we show below, they imply $\psie=0$, $\che=0$ and $\rhe=0$
for $t'\geq t$ so that the effective action reduces to
\begin{eqnarray}
\label{eq:xhe}
\fl \xe=\sum_{it} \text{i}\hat{x}_i(t)\bigg[x_i(t+\Delta)-x_i(t)+\Delta\bigg(\lambda_i x_i(t) -\lhe -\Delta\sum_{t'< t}\rhe x_i(t')\bigg) \bigg]+\notag\\
-\frac{\Delta^2}{2}\sum_{itt'}\bhe\hat{x}_i(t)\hat{x}_i(t')-\frac{\Delta}{2}\sum_{it}\Sigma_{ii}\hat{x}_i(t)\hat{x}_i(t)
\end{eqnarray}
This is exactly the action for the Langevin dynamics
\begin{equation}
\frac{x_i(t+\Delta)-x_i(t)}{\Delta}= -\lambda_i x_i(t)+ \lhe +\Delta\sum_{t'}\rhe x_i(t') +\xi_i(t)+\chi_i(t)
\end{equation}
where $\bm{\chi}$ is a coloured, local Gaussian noise with
\begin{equation}
\langle\chi_i\rangle=0 \qquad \langle\chi_i(t) \chi_i(t')\rangle= \bhe
\end{equation}
Note that the covariance of this effective noise is defined exactly so that the quadratic terms in $\hat{x}_i(t)$ in $\xe$ arise from averaging over $\chi_i$ 
\begin{equation}
\text{e}^{-\Delta^2\sum_{tt'}\hat{x}_i(t)\hat{B}^{\text{eff}}_i(t,t')\hat{x}_i(t')/2}=\langle\text{e}^{\text{i}\Delta\sum_t \hat{x}_i(t)\chi_i(t)} \rangle_{\bm{\chi}}
\end{equation}

The remainder of the analysis is easier to carry out in the continuous time-limit $\Delta\to 0$. The effective equation of motion becomes 
\begin{equation}
\label{eq:eff_dyn}
\frac{d x_i(t)}{dt}= -\lambda_i x_i(t)+ \lhe +\int_{0}^t dt'\rhe x_i(t') +\xi_i(t)+\chi_i(t)
\end{equation}
which shows that $\rhe$ plays the role of a memory function. Because this dynamics is causal, it does indeed give $\muh_i=0$, $B_i(t,t')=0$ and $R_i(t,t')=0$ for $t' \geq  t$, and so our original assumptions about the order parameter values are self-consistent.

It remains to obtain the equations for the nonzero order parameters $\mu_i(t)$, $R_i(t,t')$ for $t>t'$, and $C_i(t,t')$.
For the means we have by simple averaging over the zero mean noises $\xi_i$ and $\chi_i$ 
\begin{equation}
\label{eq:eff_mu}
\frac{d \mu_i(t)}{dt}= -\lambda_i \mu_i(t) +\int_{0}^t dt'\rhe\mu_i(t') + \lhe
\end{equation}
For the responses, standard results for linear dynamics with Gaussian noise give 
\begin{equation}
\label{eq:eff_R}
\frac{\partial R_i(t,t')}{\partial t}=\frac{\partial \dot \mu_i(t)}{\partial l^{\rm eff}_i(t')}= -\lambda_i R_i(t,t') +\int_{t'}^{t} dt'' \hat{R}^{\rm eff}_i(t,t'')R_i(t'',t')+ \delta(t-t')
\end{equation}
For the correlations it makes sense to consider the connected version $\delta C_i(t',t) = C_i(t',t) - \mu_i(t')\mu_i(t)$, which obeys
\begin{equation}
\label{eq:eff_C}
 \frac{\partial \delta C_i(t,t')}{\partial t}= -\lambda_i \delta C_i(t,t') +\int_{t'}^{t} dt'' \hat{R}^{\rm eff}_i(t,t'')\delta C_i(t'',t')+\int_{0}^{t'}dt'' R_i(t',t'')\big(\hat{B}^{\rm eff}_i(t,t'')+\Sigma_{ii}\delta(t-t'') \big)
\end{equation}
These order parameters $\mu_i(t)$, $\delta C_i(t,t')$ and $R_i(t',t)$ are uniquely determined from the above equations when supplemented with initial values $\mu_i(0)$ and $\delta C_i(0,0)$, which we assume are given as part of the specification of our system.

\subsection{First order: Mean Field equations}
\label{sec:MF}

As explained above the equilibrium case, see equations (\ref{eq:G1_first_version}) and (\ref{eq:G1_second_version}), the first order correction in $\alpha$ to the Plefka free energy is
\begin{equation}
G^1=\frac{\partial G_{\alpha}}{\partial \alpha}\bigg\vert_{\alpha=0}=\bigg\langle \frac{\partial\Xi_{\alpha}}{d \alpha} \bigg\rangle_{0}
\end{equation}
or explicitly
\begin{equation}
\label{g1}
G^1=-\Delta\sum_{it} \big \langle\text{i}\hat{x}_i(t)\phi_i(\bm{x}(t))\big\rangle_{0}
\end{equation}
For the sake of brevity we drop the subscript $0$: all averages below are to be taken at $\alpha=0$ unless otherwise specified. To find $G^1$ explicitly, consider first a generic vector $\bm{z}=\lbrace z_a \rbrace$ of Gaussian variables
with mean $\bm{\mu}$ and covariance matrix $\bm{\Gamma}$. Then by
integration by parts
\be
\langle \delta z_a \phi(\bm{z})\rangle = \sum_b \Gamma_{ab}\langle \partial_{z_b}\phi(\bm{z})\rangle
\label{eq:av_z_phi}
\ee
where $\delta z_a = z_a - \mu_a$. 
Applying this first identity to our case gives
\be
\langle \delta\hat{x}_i(t) \phi_i(\bm{x}(t))\rangle = \text{i}\delta R_i(t,t)\bigg\langle \frac{\partial \phi_i(\bm{x}(t))}{\partial x_i(t)}
\bigg\rangle
\ee
where $\delta\hat{x}_i = \hat{x}_i-\hat{\mu}_i$ and $\delta R_i(t,t')=  R_i(t,t') + \text{i} \hat{\mu}_i(t')\mu_i(t))$ is the connected response function.
As a consequence,
\begin{eqnarray}
G^1=-\Delta\sum_{it}\bigg( \text{i} \hat{\mu}_i(t)\langle\phi_i(\bm{x}(t))\rangle-\delta R_i(t,t)\bigg\langle \frac{\partial \phi_i(\bm{x}(t))}{\partial x_i(t)}\bigg\rangle\bigg)
\label{eq:G1_full}
\end{eqnarray}
While not fully explicit, the value of this expression is fully determined by our order parameters; specifically the averages over $\bm{x}(t)$ are over independent Gaussian variables $x_i(t)$ with mean $\mu_i(t)$ and variance $\delta C_i(t,t) = C_i(t,t)-\mu_i^2(t)$.

We can now obtain the first order (in $\alpha$) conjugate fields, which are the negative derivatives of $G^1$ w.r.t.\ the order parameters
\be
\bm{h}^1=-\frac{1}{\Delta^n}\frac{\partial G^1}{\partial\bm{m}},\qquad \bm{h}^1 = \lbrace \psi_i^1(t),l_i^1(t),\hat{C}_i^1(t,t'),\hat{R}_i^1(t',t),\hat{B}_i^1(t,t') \rbrace
\label{eq:h1_generic}
\ee
where according to our convention in the construction of $\Xi_\alpha$, the exponent $n=1$ for linear order parameters and $n=2$ for quadratic ones. Explicitly we obtain
\begin{subequations}
\label{eq:first_fields}
\begin{align}
\begin{split}
\psi_i^1(t)=&\quad\sum_{j} \bigg(\text{i}\hat{\mu}_{j}(t)
\frac{\partial \langle\phi_j(\bm{x}(t))\rangle}{\partial \mu_i(t)}
-\delta R_j(t,t) \frac{\partial}{\partial \mu_i(t)}\bigg\langle \frac{\partial \phi_j(\bm{x}(t))}{\partial x_j(t)}\bigg\rangle\bigg)+\\
&\quad -\text{i}\hat{\mu}_i(t)\bigg\langle \frac{\partial \phi_i(\bm{x}(t))}{\partial x_i(t)}\bigg\rangle
\end{split}\\
l_i^1(t)=&\quad\mu_i(t)\bigg\langle\frac{\partial \phi_i(\bm{x}(t))}{\partial x_i(t)}\bigg\rangle-\langle\phi_i(\bm{x}(t))\rangle\label{eq:first_fields_l}\\
\hat{C}_i^1(t,t')=&\quad \frac{1}{\Delta}\sum_{j}\bigg(\text{i}\hat{\mu}_j(t)\frac{\partial\langle \phi_j(\bm{x}(t))\rangle}{\partial C_i(t,t)}-
\delta R_j(t,t)\frac{\partial}{\partial C_i(t,t)}\bigg\langle \frac{\partial \phi_j(\bm{x}(t))}{\partial x_j(t)}\bigg\rangle\bigg)\delta_{tt'}\\
\hat{R}_i^1(t,t') =&{}-\frac{1}{\Delta}\bigg\langle\frac{\partial \phi_i(\bm{x}(t))}{\partial x_i(t)} \bigg\rangle\delta_{tt'}\label{eq:first_fields_r}\\
\hat{B}_i^1(t,t') =&\quad0 \label{eq:first_fields_b}
\end{align}
\end{subequations}
Using the general identity for Gaussian variables $\bm{z}=\lbrace z_a \rbrace$ with means $\mu_a$
\be 
\partial_{\mu_a}\langle \phi(\bm{z})\rangle = \langle \partial_{z_a}\phi(\bm{z})\rangle
\ee 
the first average in the expression for $\psi_i^1$ could also be written as $\langle \partial \phi_j(\bm{x}(t)) / \partial x_i(t)\rangle$.

The effective fields defining the effective non-interacting dynamics are now $\he = - \alpha \bm{h}^1$. To evaluate these we can exploit that the final order parameter values should obey $\hat\mu_i(t)=0$ and $R_i(t,t)=0$, hence also $\delta R_i(t,t)=0$. This then gives $\psi_i^1(t)=0$ and $\hat{C}_i^1(t,t)=0$ so that also the corresponding effective fields vanish, as anticipated above in our general discussion of the effective non-interacting dynamics. Note that it is important to make the above simplifying assumptions only in the final expressions for the effective fields, not already in $G^1$ as derivatives w.r.t.\ e.g.\ $\hat\mu_i(t)$ do contribute to the effective fields.

The only remaining nonzero effective fields at this stage are $\lhe=-\alpha l^1_i(t)$ and $\rhe=-\alpha \hat{R}^1_i(t,t')$. We insert these into \eqref{eq:xhe} to get the mean field equations for the now effectively non-interacting degrees of freedom $x_i(t)$
\begin{equation}
\frac{d x_i(t)}{dt} = -\lambda_i x_i(t)+\alpha\bigg\langle\frac{\partial \phi_i(\bm{x}(t))}{\partial x_i(t)} \bigg\rangle(x_i(t)-\mu_i(t)) +\alpha\langle\phi_i(\bm{x}(t))\rangle+ \xi_i(t)  
\end{equation}
Not unexpectedly for an effective linear dynamics, the interaction term $\phi_i(\bm{x}(t))$ has here effectively been linearized in deviations of $x_i(t)$ from its mean. The self-consistency equation for this mean reads
\begin{equation}
\frac{d\mu_i(t)}{dt} =  -\lambda_i \mu_i(t) +\alpha\langle\phi_i(\bm{x}(t))\rangle
\end{equation}
The equations for the equal-time correlations $C_i(t,t)$ can be obtained from the equation of motion for the fluctuations around the mean $\dx_i(t)=x_i(t)-\mu_i(t)$
\bea
\frac{d \dx_i(t)}{dt}  &=& \bigg(-\lambda_i +\alpha\bigg\langle\frac{\partial \phi_i(\bm{x}(t))}{\partial x_i(t)} \bigg\rangle\bigg) \dx_i(t)
 + \xi_i(t)
 \label{eq:dxi_dt_first_order}
\eea
This gives directly, in the standard manner for an Ornstein-Uhlenbeck process with time-dependent drift,
\begin{equation}
\frac{d\delta C_i(t,t)}{dt} = 2\bigg(-\lambda_i +\alpha\bigg\langle\frac{\partial \phi_i(\bm{x}(t))}{\partial x_i(t)} \bigg\rangle\bigg)\delta C_i(t,t)+\Sigma_{ii}
\end{equation}
In general, the above equations need to be solved jointly for the $2N$ time-dependent order parameters $\mu_i(t)$ and $C_i(t,t)$; this is because the average of $\partial \phi_i / \partial x_i$ generically depends on both means and variances. The case of purely linear interactions, where $\phi_i = \sum_{j\neq i} K_{ij} x_j$, is an obvious exception: here the equations for the means do not involve the variances so can be solved separately.

It is worth commenting at this stage how our first order result compares with that of a conventional Plefka approach that constrains only the first moments $\mu_i(t)$ and $\hat\mu_i(t)$. 
The effective field terms in the effective dynamical action are then linear in $x_i(t)$ and $\hat x_i(t)$. This means that all second order fluctuation statistics remain as in a non-interacting problem.
In particular, $\delta C_i(t,t')$ and $R_i(t,t')$ do not feel any effect of the non-trivial drift $\phi_i$. The second term in the brackets in the r.h.s.\ of \eqref{eq:dxi_dt_first_order} would be absent, and the interaction
term $\phi_i$ would only appear via its average. Already to first order in $\alpha$ it is clear, then, that the extended Plefka approach captures qualitatively more of the dynamics of the interacting system than a
conventional Plefka method constraining linear averages.

\subsection{Second order: TAP equations}
\label{sec:TAP}
The second order of the Plefka free energy can be evaluated starting from the equality \eqref{eq:roudi}
\begin{equation}
G^2=\frac{\partial^2 G_{\alpha}}{\partial \alpha^2}\bigg\vert_{\alpha=0}=\bigg \langle  \bigg(\delta\,\frac{d \Xi_{\alpha}}{d\alpha}\bigg)^2 \bigg \rangle_{0} 
\end{equation}
Including the first order fields in the effective action, with the prefactor
$\alpha$, and taking $\frac{d \Xi_{\alpha}}{d \alpha}$ at $\alpha=0$ gives

\begin{eqnarray}
\label{eq:dXi_alpha_TAP}
\fl\frac{d\Xi_{\alpha}}{d \alpha}\bigg\vert_{\alpha=0}= -\Delta\sum_{it}\text{i}\hat{x}_i(t)\phi_i(\bm{x}(t))+ \Delta\sum_{it}\psi^1_i(t)\bigg(x_i(t)-\mu_i(t)\bigg)-\Delta\sum_{it}l_i^1(t)\bigg(\text{i}\hat{x}_i(t)-\text{i}\hat{\mu}_i(t)\bigg)+\notag \\ 
\fl+\Delta^2\sum_{itt'}\hat{C}^1_i(t,t')\bigg(x_i(t)x_i(t')-C_i(t,t')\bigg)+\Delta^2\sum_{itt'}\hat{R}^1_i(t,t')\bigg(-\text{i}\hat{x}_i(t)x_i(t')-R_i(t',t)\bigg)
\end{eqnarray}

While the following analysis can be carried out for general drift $\phi_i(\bm{x})$, we will restrict the scenario slightly by assuming that
\be
\frac{\partial \phi_i(\bm{x})}{\partial x_i}=0
\ee
as this significantly reduces the number of terms in the expressions. Intuitively, we are assuming that 
$\phi_i(\bm{x})$ is a function only of the other variables $x_j$; equivalently, $x_i$ interacts with itself only via the linear term $-\lambda_i x_i$. In the later steps of the calculation, from \eqref{secondh},
we will add the assumption that the drift $\phi_i$ is an additive combinations of functions of the other variables $x_j$, i.e.\ of the form $\phi_i(\bm{x})=\sum_{j\neq i}g_{ij}(x_j)$.
The above expressions for the first order conjugate fields then simplify to
\begin{subequations}
\begin{align}
\psi_i^1(t)=&\quad\sum_{j}\text{i}\hat{\mu}_{j}(t)\frac{\partial \langle\phi_j(\bm{x}(t))\rangle}{\partial \mu_i(t)}\\
l_i^1(t)=&-\langle\phi_i(\bm{x}(t))\rangle\\
\hat{C}_i^1(t,t')=&\quad \frac{1}{\Delta}\sum_{j}\text{i}\hat{\mu}_j(t)\frac{\partial\langle \phi_j(\bm{x}(t))\rangle}{\partial C_i(t,t')}\delta_{tt'}\\
\hat{R}_i^1(t,t') =&\quad0\\
\hat{B}_i^1(t,t') =&\quad0
\end{align}
\end{subequations}
Inserting these into \eqref{eq:dXi_alpha_TAP} one finds
\begin{eqnarray}
\fl\delta\,\frac{d \Xi_{\alpha}}{d \alpha}=\frac{d \Xi_{\alpha}}{d \alpha} - \bigg\langle\frac{d \Xi_{\alpha}}{d \alpha} \bigg\rangle_{0}=&-\Delta\sum_{it} \bigg[ \text{i} \delta \hat{x}_i(t)\delta\phi_i(\bm{x}(t))+\text{i} \hat{\mu}_i(t)\sum_{j}\bigg(\delta \phi_i(\bm{x}(t))+\notag\\
&-\frac{\partial \langle \phi_i(\bm{x}(t))\rangle}{\partial \mu_j(t)}\delta x_j(t) - \frac{\partial \langle \phi_i(\bm{x}(t))\rangle}{\partial C_j(t,t)}\delta (x_j(t)x_j(t))    \bigg) \bigg]
\end{eqnarray}
where $\delta \hat{x}_i(t)=\hat{x}_i(t)-\hat{\mu}_i(t)$ as before and $\delta  \phi_i(\bm{x}(t))= \phi_i(\bm{x}(t))-\langle \phi_i(\bm{x}(t)) \rangle$, while $\delta(x_j(t)x_j(t)) = x_j^2(t) - C_j(t,t)$. 
To calculate $G^2$ one now needs to square this and evaluate the relevant averages, expressing them in terms of the relevant order parameters \eqref{eq:moments}. Because the averages are taken at $\alpha=0$, there are no correlations between variables at different sites $i$.
For the same reason all statistics are Gaussian, and one can use Wick's theorem to reduce all higher order moments to first and second order ones.

Once $G^2$ has been found, the $O(\alpha^2)$ corrections for the fields can be calculated from
\be
\label{secondh}
\bm{h}^2=-\frac{1}{\Delta^n}\frac{\partial G^2}{\partial\bm{m}}
\ee
which is just the second order analogue of \eqref{eq:h1_generic}. With these general expressions for the fields we obtained, one can again impose the physical constraints on the order parameters, i.e.\ $\muh_i(t)=0$, $\dR_i(t,t')=0$ for $t'\geq t$ and $\delta B_i(t,t')=B_{i}(t,t') + \hat{\mu}_i(t)\hat{\mu}_i(t') =0$. We omit the details and write directly the final simplified form of the second order fields 
\begin{subequations} 
\begin{align}
\psi^2_i(t)&= 0\\
l^2_i(t)&=2\Delta\sum_{jt'} \bigg\langle \frac{\partial \phi_{i}(\bm{x}(t))}{\partial x_j(t)}\, \frac{\partial \phi_{j}(\bm{x}(t'))}{\partial x_i(t')}  \bigg\rangle\mu_i(t')  R_j(t,t')\\
\hat{R}^2_i(t,t') &=-2\sum_{j}\bigg\langle \frac{\partial \phi_{i}(\bm{x}(t))}{\partial x_j(t)}\, \frac{\partial \phi_{j}(\bm{x}(t'))}{\partial x_i(t')}  \bigg\rangle  R_j(t,t')\\
\hat{C}^2_i(t,t') &= 0\\
\hat{B}^2_i(t,t') &=-\big\langle \delta \phi_{i}(\bm{x}(t))\delta \phi_{i}(\bm{x}(t')) \big\rangle
\end{align}
\end{subequations}
These fields, multiplied by $-\frac{\alpha^2}{2}$, give the second order contributions to the effective fields in the non-interacting dynamical action, $\he=-\alpha \bm{h}^1-\frac{\alpha^2}{2}\bm{h}^2$. One sees that $\psie$ and $\che$ remain identically null also to second order, while the nonzero effective fields are, in the continuous time limit $\Delta\to 0$
\begin{equation}
\begin{split}
\lhe&= \alpha\big\langle \phi_{i}(\bm{x}(t)) \big\rangle- \alpha^2\int_{0}^t dt'\sum_{j}\bigg\langle \frac{\partial \phi_{i}(\bm{x}(t))}{\partial x_j(t)} \,\frac{\partial \phi_{j}(\bm{x}(t'))}{\partial x_i(t')}  \bigg\rangle\mu_i(t')R_j(t,t')
\end{split}
\end{equation}
\begin{equation}
 \rhe=\alpha^2\sum_{j}\bigg\langle \frac{\partial \phi_{i}(\bm{x}(t))}{\partial x_j(t)} \,\frac{\partial \phi_{j}(\bm{x}(t'))}{\partial x_i(t')}  \bigg\rangle R_j(t,t')
\end{equation}
\begin{equation}
\bhe=\alpha^2\big\langle \delta \phi_{i}(\bm{x}(t))\delta \phi_{i}(\bm{x}(t')) \big\rangle
\end{equation}
We notice that the causality structure of $\rhe$ is directly related to that of $R_i(t,t')$, i.e.\ both are nonzero only when the second time argument is smaller than the first. (In the first order calculation we had in addition found a nonzero equal-time value for $\rhe$ but this was due to a self-interaction that we have since assumed to be zero.)
Substituting the fields into $\xe$ \eqref{eq:xhe}, we obtain the uncoupled description of the dynamics to second order in $\alpha$
\begin{equation}
\frac{d x_i(t)}{dt} =  -\lambda_i x_i(t)+\alpha\big\langle \phi_{i}(\bm{x}(t)) \big\rangle+\alpha^2\sum_{j}\int_{0}^t dt' \bigg\langle \frac{\partial \phi_{i}(\bm{x}(t))}{\partial x_j(t)} 
\,\frac{\partial \phi_{j}(\bm{x}(t'))}{\partial x_i(t')}  \bigg\rangle R_j(t,t')\delta x_i(t')+\xi_i(t)+\chi_i(t)
\label{eq:dx_dt_TAP}
\end{equation}
The dynamical TAP equations are then the self-consistency equations for the $\mu_i(t)$, $R_i(t,t')$ and $C_i(t,t')$ that result. These are written in their general form in \eqref{eq:eff_mu} to \eqref{eq:eff_C} above. What is remarkable is that the integral over the past in \eqref{eq:dx_dt_TAP} does not contribute to the evolution equation for the means, which as to first order is given by
\begin{equation}
\frac{d \mu_i(t)}{dt}=-\lambda_i\mu_i(t)+\alpha\big\langle \phi_{i}(\bm{x}(t)) \big\rangle
\end{equation}
This does not mean, of course, that the actual time courses $\mu_i(t)$ wil be the same in the MF and TAP equations: the TAP equations for the variances $C_i(t,t)$ are different from MF, and these variances affect the average $\langle \phi_i\rangle$ in the evolution of the $\mu_i(t)$.

\subsection{Linear case}
\label{sec:LD}
It is instructive to consider this framework for a simple case, i.e. a differential equation with linear couplings
\begin{equation}
\label{eq:LDeq}
\frac{dx_i(t)}{dt} = - \lambda_{i}x_i(t)+\sum_{j} K_{ij}x_j(t) +\xi_i(t)
\end{equation}
This corresponds to the choice $\phi_i(\bm{x}) = \sum_{j} K_{ij} x_j$ for the drift. We assume throughout that $K_{ii}=0$, so that there is no self-interaction in $\phi_i$. 

\subsubsection{First order: Mean Field}

The first order in $\alpha$ of the Plefka free energy $G$ simplifies from \eqref{g1} to
\begin{equation}
\begin{split}
G^1=-\Delta \sum_{it}\text{i}\hat{\mu}_i(t)\sum_{j} K_{ij}\mu_j(t)
\end{split}
\end{equation}
and gives the first order fields
\begin{subequations}
\begin{align}
\psi_i^1(t)&= \sum_{j} \text{i}\hat{\mu}_{j}(t)K_{ji}\\
l_i^1(t)&= -\sum_jK_{ij}\mu_j(t)\\
\hat{R}_i^1(t,t')& =\hat{C}^1_i(t,t')=\hat{B}^1_i(t,t')=0
\end{align}
\end{subequations}
The effective dynamical equation becomes
\begin{equation}
\frac{d x_i(t)}{dt} = - \lambda_i x_i(t)+\alpha \sum_{j} K_{ij}\mu_j(t)+\xi_i(t)  
\end{equation}
and gives for the means the equations of motion
\begin{equation}
\frac{d\mu_i(t)}{dt} =  - \lambda_i \mu_i(t)+\alpha\sum_{j} K_{ij}\mu_j(t)
\end{equation}
For $\alpha=1$ these agree with the exact equations. 
The second order fluctuation statistics, on the other hand, are  unchanged from the non-interacting system at this level of approximation.

\subsubsection{Second order: TAP}
The effective dynamics to second order in $\alpha$ become, as a special case of \eqref{eq:dx_dt_TAP}
\begin{equation}
\label{eq:dx_dt_TAP1}
\frac{d x_i(t)}{dt} =- \lambda_i x_i(t)+  \alpha \sum_{j} K_{ij}\mu_j(t)  + \alpha^2 \int_{0}^{t}dt'\sum_{j}K_{ij}R_j(t,t')K_{ji}\delta x_i(t')+\xi_i(t)+\chi_i(t)
\end{equation}
In the integral term we have arranged the factors to allow a simple intuitive interpretation: a fluctuation $\delta x_i$ at time $t'$ acts via $K_{ji}$ as an effective field on $x_j$; at time $t$ this produces a response in $x_j$ modulated by $R_j(t,t')$, which then acts back on $x_i$ via $K_{ij}$.

Putting $\alpha=1$, the mean dynamics is identical to the (already exact) MF description
\begin{equation}
\label{eq:mean2o}
\frac{d\mu_i(t)}{dt} = - \lambda_i \mu_i(t)+\sum_{j}K_{ij}\mu_j(t)
\end{equation}
Responses have their temporal evolution governed by \eqref{eq:eff_R}
\begin{equation}
\label{resplin}
\frac{\partial R_i(t,t')}{\partial t}=  - \lambda_i R_i(t,t') +\sum_j\int^t_{t'} dt''K_{ij} R_j(t,t'') K_{ji} R_i(t'',t') +\delta(t-t')
\end{equation}
while for the connected correlations one has, from \eqref{eq:eff_C}
\begin{eqnarray}
\label{corrlin}
\frac{\partial\,\delta C_i(t,t')}{\partial t}& =  - \lambda_i \delta C_i(t,t') +\sum_{j}\int_{0}^t dt''K_{ij} R_j(t,t'') K_{ji}\delta C_i(t'',t') +\notag\\
&+\Sigma_{ii}R_i(t',t) + \sum_{j}\int_{0}^{t'}dt''R_i(t',t'')K_{ij}^2\delta C_j(t,t'') 
\end{eqnarray}
The last term involves the covariance of the coloured noise $\chi_i(t)$, which is $\sum_j K_{ij}^2 \delta C_j(t,t')$.

\section{Exactness in the thermodynamic limit}
\subsection{Motivation and setup}
\label{sec:LD1}

The extended Plefka expansion derived above is, of course, an approximation in general because we have truncated the power series expansion in the interaction strength $\alpha$ at second order. We would
expect the approximation to become exact, however, provided that the interactions between variables are suitably long-ranged and we take the thermodynamic limit $N\to\infty$ of a large system: a central
limit theorem argument then suggests that the interactions have Gaussian statistics as the extended Plefka expansion predicts. The purpose of this section is to study in detail one example of a model 
in this class, namely the linear interaction model introduced in section \ref{sec:LD} with random couplings $K_{ij}$. There are rather more general scenarios where we expect our method to give the exact
results, as discussed in section \ref{sec:conclusions} below.

We already know (see \ref{eq:eff_mu}) that the extended Plefka equations for the means are exact, and will show that the responses and correlations are also predicted correctly by the extended Plefka approach. The exact solution that we work out as our baseline has close similarities with the analysis of the $p=2$-spin spherical model;  see \cite{cugliandolo} for a detailed study of the latter.

We will focus on the long-time limit $t\to\infty$, where the analysis simplifies because two-time correlations and responses become time translation invariant (TTI), i.e.\ depend only on time differences. The derivation of the extended Plefka expansion does not of course rely on TTI, and we would expect that the agreement with the exact solution can be demonstrated also for transient relaxation to the steady state.

To be specific, we consider the linear dynamics \eqref{eq:LDeq}; this corresponds to the Langevin dynamics of a $p=2$-spin spherical model where the spins are replaced by arbitrary degrees of freedom $x_i(t)$ interacting in pairs.
For the sake of simplicity we assume $\lambda_i=\lambda$ and $\Sigma_{ii}=\Sigma$ for all $i=1,\ldots,N$, i.e.\ we take the self-interaction and noise strength as the same for all degrees of freedom.
The self-interaction plays the role of the Lagrange multiplier enforcing the spherical constraint in the $p=2$-spin spherical model: note that in our case it is not time dependent, however, but simply a constant.

We want to proceed with as few restrictive assumptions on the couplings $K_{ij}$ as possible; in fact, nothing in the derivation of the Plefka expansion requires particular conditions on $\bm{K}$. 
A simple choice is then to suppose that $\bm{K}$ is a real matrix with elements that are randomly distributed Gaussian variables with zero mean and variance $\langle K_{ij}^2\rangle=1/N$, drawn independently except for the correlation
\begin{equation}
\label{kji}
 \langle {K}_{ji} {K}_{ij} \rangle=\frac{\eta}{N}
\end{equation}
The parameter $\eta\in[-1,1]$ controls the degree to which the matrix $\bm{K}$ is symmetric, i.e.\ it is a measure of symmetry for the physical couplings in the system. Such ensembles of matrices with Gaussian-distributed elements were first studied by Girko \cite{girko} and Ginibre \cite{ginibre}: their characteristic feature is that unless $\eta=1$,
the eigenvalues are not restricted to the real axis but distributed over an area in the complex plane.

For $\eta=1$ we have symmetric matrices, which belong to what is known as the Wigner or Gaussian Orthogonal Ensemble. Symmetry here ensures that the dynamics obeys detailed balance with respect to the energy function $\sum_i \lambda x_i^2/2 + \sum_{ij}x_iK_{ij}x_j/2$ so that the stationary regime is an equilibrium state.

The value $\eta=0$ means that all correlations between matrix elements vanish and thus identifies a fully asymmetric $\bm{K}$: such random matrices, with completely independent real entries, belong to the Ginibre Orthogonal Ensembles~\cite{ginibre}. Finally, $\eta=-1$ describes the antisymmetric case, where all eigenvalues of $\bm{K}$ lie along the imaginary axis because $\text{i}\bm{K}$ is Hermitian.

\subsection{Extended Plefka expansion}
\label{sec:TLLD}

We next evaluate the predictions of the extended Plefka approach for our system with equation of motion \eqref{eq:LDeq}. 
As shown in \eqref{eq:dx_dt_TAP1} above, the effective single-site dynamics is given by
\begin{equation}
\label{eq:pl}
\frac{d x_i(t)}{dt}= - \lambda x_i(t)+ \sum_{j} K_{ij}\mu_j(t) +\int_0^t dt' \sum_{j}K_{ij}R_j(t,t')K_{ji}\delta x_i(t')+\phi_i(t)
\end{equation}
where 
\begin{equation}
\label{pl1}
\phi_i(t)=\xi_i(t)+\chi_i(t) \qquad \langle\phi_i(t) \phi_i(t')\rangle=\Sigma\delta(t-t') + \sum_{j}K_{ij}^2 \delta C_j(t, t')
\end{equation}
The dynamics of the means $\mu_i(t)$, obtained by averaging over the ensemble \eqref{eq:pl} as in \eqref{eq:mean2o}, is in full agreement with the exact one obtained by simply taking the mean of \eqref{eq:LDeq}.

Let us calculate the response, which in the Plefka approach is given by \eqref{resplin}. 
The dependence on the site $i$ on the r.h.s.\ arises only from the term $\sum_j K_{ij}K_{ji}R_j(t,t')$. Because $K_{ij}K_{ji}$ is of order $1/N$, while the $R_j$ are of order unity and are expected to have vanishing correlation 
with $K_{ij}K_{ji}$ 
(for any fixed $i$) for large $N$, this sum is self-averaging: for large $N$ it can be replaced by 
\begin{displaymath}
 \sum_{j}K_{ij}K_{ji}R_j(t,t')\sim \frac{\eta}{N}\sum_{j}R_j(t,t')\equiv \eta R(t,t')
\end{displaymath}
because $\langle K_{ij}K_{ji}\rangle = \eta/N$. For later we note that the non-trivial term in the noise covariance \eqref{pl1} self-averages similarly to
\begin{displaymath}
\sum_{j}K_{ij}^2 C_j(t, t')\sim \frac{1}{N}\sum_{j} C_j(t, t')\equiv C(t,t')
\label{eq:C_self_averaging}
\end{displaymath}
The self-averaged version of \eqref{resplin} now reads
\begin{equation}
\frac{\partial R_i(t,t')}{\partial t}=  - \lambda R_i(t,t') +\eta\int_{t'}^{t} dt''R(t,t'')R_i(t'',t') +\delta(t-t')
\end{equation}
From this one sees that all sites $i$ will have the same response for large $N$, which makes sense because with our 
long-range disordered couplings all sites $i$ become equivalent. We can thus drop the site index on $R_i$ from now on, or formally average over $i$ to get an equation for $R$. 

As explained above we now consider the long-time limit where a steady state should be reached so that the response becomes TTI, $R(t,t')=R(t-t')$
\begin{equation}
\label{RePl}
\frac{\partial R(t-t')}{\partial t}= - \lambda R(t-t') + \eta\int_{t'}^{t} dt''R(t-t'')R(t''-t')+\delta(t-t')
\end{equation}
Laplace transforming with respect to time differences, with $z$ the conjugate variable, gives
\begin{equation}
 (z+\lambda)\tilde{R}(z)= \eta\tilde{R}^2(z)+1
 \label{eq:Rz_Plefka}
\end{equation}
This second order equation for the Laplace transformed response $\tilde{R}(z)$ has solution
\begin{equation}
 \label{eq:PleRes}
 \tilde{R}(z)= \frac{1}{2\eta}(z+\lambda)-\frac{1}{2\eta}\sqrt{(z+\lambda)^2-4\eta}
\end{equation}
Here the sign is chosen to retrieve the correct behaviour for $z\to \infty$: as $R(t-t')$ must approach unity for small time differences, the Laplace transform $\tilde{R}(z)$ has 
to decay as $1/z$ for large $z$. The result \eqref{eq:PleRes} is particularly simple for $\eta=0$, where the response takes the form
\begin{equation}
\label{reAsPl}
 \tilde{R}(z)=\frac{1}{z+\lambda}
\end{equation}

We next apply the same approach to the calculation of the connected correlations $\delta C(t,t')$. As we will only consider connected correlations in the following we drop the $\delta$ and write simply $C(t,t')$. We start from \eqref{corrlin}, make the self-averaging replacement \eqref{eq:C_self_averaging}, drop the site index and obtain 
\begin{equation}
\label{CoPl}
 \frac{\partial C(t-t')}{\partial t}= -\lambda  C(t-t') + \eta\int_{-\infty}^{t} dt'' R(t-t'')C(t''- t')+ \int_{-\infty}^{t'} dt''\big[\Sigma \delta(t-t'')+C(t- t'')\big]R(t'- t'')
\end{equation}
We take a two-sided Laplace transform of this
\begin{equation}
z\tilde{C}(z) = -\tilde{C}(z) + \eta\tilde{R}(z)\tilde{C}(z)
+ [\Sigma + \tilde{C}(z)]\tilde{R}(-z)
\end{equation}
and solve to get
\begin{equation}
\label{eq:corple}
 \tilde{C}(z)=\frac{\Sigma\tilde{R}(-z)}{z+\lambda-\tilde{R}(-z)-\eta\tilde{R}(z)}=\frac{\Sigma\tilde{R}(z)\tilde{R}(-z)}{1-\tilde{R}(z)\tilde{R}(-z)}
\end{equation}
In the second equality we have simplified using \eqref{eq:Rz_Plefka} to obtain a form that is manifestly even in $z$, as it should be because $C(t-t')=C(t'-t)$. Note that for the response, which is causal so vanishes for negative time differences, the two-sided Laplace transform reduces to the one-sided version.

 \subsection{Exact Solution}
 \label{sec:ED}
To assess whether the above predictions of the extended Plefka method are correct, we now study the exact solution of our model.  

We will require as an essential ingredient the spectral density $\rho(k)$ of $\bm{K}$ in the thermodynamic limit, which follows from general theorems, namely Girko's elliptic and circular laws and the Wigner semicircular law. Girko's
elliptic law \cite{girko1} states that the average eigenvalue distribution $\rho(k)$ of $N\times N$ random matrices $\bm{K}$ drawn from a Gaussian ensemble described by
 \eqref{kji}, in the limit $N\rightarrow\infty$, is
\begin{equation}
  \rho(k)=\begin{cases}
    \frac{1}{\pi(1-\eta^2)} & \left(\frac{x}{1+\eta}\right)^2+\left(\frac{y}{1-\eta}\right)^2< 1\\
    0 & \text{otherwise}
  \end{cases}
\end{equation}
where we have written $x$ and $y$ for the real and imaginary values of the eigenvalue $k$. The density $\rho(k)$ is uniform in an ellipse in the complex plane whose semi-axes are $1+\eta$ and $1-\eta$, respectively, along the real and imaginary directions, and whose foci are $\pm 2 \sqrt{\eta}$. 
In the limit $\eta\rightarrow 1$ the Wigner semicircle law \cite{mehta} is recovered from this, for the distribution of real eigenvalues of matrices from the Wigner ensemble
\begin{equation}
\label{eq:proj}
 \rho(k)= \frac{1}{2\pi}\sqrt{4-k^2} \qquad k\in [-2, 2]
\end{equation}
Girko's elliptic law can then be regarded as the generalization of Wigner's semicircular law to the case of an arbitrary degree of symmetry. For $\eta=0$ the ellipse 
degenerates into the unit circle. Let us consider the vectorial form of the dynamics \eqref{eq:LDeq} of our model, where we temporarily add an external field $\bm{l}$ on the r.h.s.
\begin{equation}
\label{eq:LDeqvec}
 \frac{d \bm{x}(t)}{dt}=-\lambda \bm{x}(t)+\bm{K}\bm{x}(t)+\bm{\xi}(t) + \bm{l}(t)
\end{equation}
The solution can be written symbolically as, if we ignore contributions from the initial conditions
\begin{equation}
 \bm{x}(t)=\int_{0}^t dt' \text{e}^{(-\lambda+\bm{K})(t-t')}[\bm{\xi}(t') + \bm{l}(t')]
 \label{eq:gensol}
\end{equation}
This gives directly for the response function matrix
\begin{equation}
\bm{R}(t,t')= \frac{\partial \langle \bm{x}(t)\rangle}{\partial \bm{l}(t')}\bigg\vert_{\bm{l}=0}= \theta(t-t') \text{e}^{(-\lambda+\bm{K})(t-t')}
\end{equation}
and we can set the field to zero again from now on. One sees that $\lambda$ must be greater than the real part of all eigenvalues $k$ of $\bm{K}$, to avoid exponentially increasing solutions. 
As the expression for $\bm{R}$ is TTI, it has a simple representation in the Laplace domain
\begin{equation}
\label{eq:LapRes}
\tilde{\bm{R}}(z)=\int_0^{+\infty}\,  \text{e}^{(\bm{K}-\lambda)s} \text{e}^{-z s} d s= [z-(\bm{K}-\lambda)]^{-1}\qquad s \equiv t-t'
\end{equation}

For comparison with the Plefka approach we are interested in $R(t-t')=(1/N)\sum_i R_{ii}(t-t')=\text{Tr}\,\bm{R}(t-t')$ if we denote by Tr the normalized trace.
This can be evaluated by integrating over the spectral density
\begin{equation}
\label{eq:intresp}
\tilde{R}(z)= \langle \text{Tr}\,\tilde{\bm{R}}(z)\rangle=\int dk\,\rho(k)  [z-(k-\lambda)]^{-1}=\begin{cases}
    \frac{1}{2\eta}(\lambda+z)-\frac{1}{2\eta}\sqrt{(\lambda+z)^2-4\eta} & \eta \ \ \text{generic}\\
    \frac{1}{2}(\lambda+z)-\frac{1}{2}\sqrt{(\lambda+z)^2-4}& \eta=1\\
    \frac{1}{z+\lambda}& \eta=0
  \end{cases} 
\end{equation}
Note that the expressions \eqref{eq:intresp} are valid only for $z+\lambda$ outside the support of the eigenvalue spectrum as otherwise the integrand has singularities. Meaningful values can still be assigned to the integral for $z$ inside the support, by appropriate regularization, and this is necessary when $\tilde{R}(z)$ is regarded as a resolvent from which spectral information is to be obtained, see e.g. \cite{chalker} and \cite{crisanti} for an interesting analogy with a two-dimensional classical electrostatic field calculation. In our case $\tilde{R}(z)$ is a Laplace transform, as it was in the Plefka calculation, so we are only interested in its behaviour for large enough real $z$ and the analytic continuation from this region, which is exactly what
\eqref{eq:intresp} provides. To be precise, \eqref{eq:intresp} with the square root assigned its principal value is valid for ${\rm Re}(z)>-\lambda$, i.e.\ to the right of the 
midpoint of the branch cut between $z= - \lambda-2\sqrt{\eta}$ and $z= - \lambda+2\sqrt{\eta}$; to the left, one has to use the negative of the principal value to ensure that 
$\tilde{R}(z)$ is analytic except in the branch cut.
Comparing with \eqref{eq:PleRes}, we thus conclude that the extended Plefka method gives the exact response function for our system.

We next turn to the correlation function. To obtain the exact expressions for this we have to resort to different tools. Information about the spectrum is no longer enough, we also require 
the statistics of correlations between the left and right eigenvectors of $\bm{K}$; these eigenvectors are different in the generic case where $\bm{K}$ is not Hermitian, i.e.\ for $\eta \neq \pm 1$.
Eigenvector statistics in non-Hermitian random matrix ensembles were studied extensively by Chalker and Mehlig \cite{chalker} and we exploit their approach, slightly adjusted for our case of matrices with real rather than complex elements.

As for the response we start from the full non-local correlation matrix, which from \eqref{eq:gensol} is given by
\begin{eqnarray}
 \fl\bm{C}(t,t')= \langle \bm{x}(t)\bm{x}^{\rm T}(t')\rangle=
\int_{0}^{t} \int_{0}^{t'} dt''dt''' \text{e}^{(-\lambda+\bm{K})(t-t'')}\langle \bm{\xi}(t'')\bm{\xi}^{\rm T}(t''')\rangle \text{e}^{(-\lambda+\bm{K}^{\rm T})(t'-t''')}\notag\\
 =\Sigma \int_0^{\text{min}(t,t')}dt'' \text{e}^{(-\lambda+\bm{K})(t-t'')}\text{e}^{(-\lambda+\bm{K}^{\rm T})(t'-t'')}
\label{eq:Cmatrix_generic_t_tt}
\end{eqnarray}
In terms of the equal-time correlator
\begin{equation}
 \bm{C}(t,t)=\Sigma \int_{0}^{t}d\tau\,\text{e}^{(-\lambda+\bm{K})\tau}\text{e}^{(-\lambda+\bm{K}^{\rm T})\tau}
\end{equation}
this simplifies to
\begin{equation}
\bm{C}(t,t')=\begin{cases}\text{e}^{(-\lambda+\bm{K})(t-t')}\bm{C}(t',t')&t\geq t'\\
\bm{C}(t,t)\text{e}^{(-\lambda+\bm{K}^{\rm T})(t'-t)} &t'>t
\end{cases}
\end{equation}
In the long-time limit $\bm{C}(t,t')$ will become TTI again, with $\bm{C}(t,t')=\bm{C}(t-t')$; $\bm{C}(0)$ then is the long-time limit of $\bm{C}(t,t)$. 
Combining the expressions for the two relative orderings of $t$ and $t'$ above and performing a two-sided Laplace transform with respect to the time difference gives
\begin{equation}
 \label{firMan}
  \tilde{\bm{C}}(z)=\bm{C}(0)(-z + \lambda - \bm{K}^{\rm T})^{-1} + (z + \lambda - \bm{K})^{-1}\bm{C}(0)
 \end{equation}
For further analysis it is useful to rewrite $\bm{C}(0)$ as
 \begin{eqnarray}
  \bm{C}(0)&=\Sigma\int_{0}^{\infty}d\tau_1\int_{0}^{\infty}d\tau_2 \,\text{e}^{(-\lambda+\bm{K})\tau_1}\text{e}^{(-\lambda+\bm{K}^{\rm T})\tau_2}\delta(\tau_1-\tau_2)=\notag\\
&=\Sigma\int_0^{\infty}d \tau_1 \int_0^{\infty} d \tau_2 \int_{-\infty}^{+\infty} \frac{d \omega}{2\pi} \, \text{e}^{\text{i} \omega (\tau_1-\tau_2)} \, \text{e}^{(-\lambda+ \bm{K})\tau_1}\,\text{e}^{(-\lambda+\bm{K}^{\rm T})\tau_2}
 \end{eqnarray}
so that after integrating over $\tau_1$ and $\tau_2$ one has
  \begin{equation}
  \label{eq:fortrace}
  \bm{C}(0)=\Sigma\int_{-\infty}^{+\infty} \frac{d \omega}{2\pi} (\lambda- \text{i}\omega - \bm{K})^{-1}(\lambda + \text{i}\omega- \bm{K}^{\rm T})^{-1}
 \end{equation}
 For a comparison with the $\tilde{C}(z)$ obtained in the Plefka approximation we need the normalized trace again, as in the case of the response, and combining \eqref{firMan} and \eqref{eq:fortrace} this takes the form
  \begin{equation}
  \label{eq:trace}
  \text{Tr}\,\tilde{\bm{C}}(z)= \Sigma\int_{-\infty}^{+\infty} \frac{d \omega}{2\pi} \, \frac{1}{z-\text{i}\omega}\bigg\lbrace\big\langle\text{Tr}\big[(\lambda-z - \bm{K})^{-1}(\lambda + \text{i}\omega - \bm{K}^{\rm T})^{-1}\big]\big\rangle-\big\langle\text{Tr}\big[(\lambda- \text{i}\omega - \bm{K})^{-1}(\lambda + z - \bm{K}^{\rm T})^{-1}\big]\big\rangle\bigg\rbrace 
  \end{equation}
where the simple matrix identity
\begin{equation}
(a-\bm{A})^{-1}-(b-\bm{A})^{-1}= (a-\bm{A})^{-1}(b-\bm{A})^{-1}(b-a)
  \end{equation}
 has been applied. We have explicitly added an average over the random sampling of $\bm{K}$ in order to be able to use random matrix technique for further evaluation. 
 This is justified because like the response, which depends only on the spectrum and is self-averaging for large $N$ because the spectrum is, 
 the correlation function is also expected to be self-averaging. For later notational convenience we have also transformed $z \rightarrow -z$ on the r.h.s.\ of \eqref{eq:trace}, anticipating 
 that the final result \eqref{eq:fincor} will be even in $z$.

The benefit of the above manipulations is that the calculation of the exact correlations is now reduced to finding the quadratic resolvents
 \begin{equation}
 \label{eq:res1}
 \big\langle \text{Tr}\big[(\lambda- \text{i}\omega - \bm{K})^{-1}(\lambda + z - \bm{K}^{\rm T})^{-1}\big]\big\rangle
 \end{equation}
 \begin{equation}
 \label{eq:res2}
\big\langle\text{Tr}\big[(\lambda-z - \bm{K})^{-1}(\lambda + \text{i}\omega - \bm{K}^{\rm T})^{-1}\big]\big\rangle
 \end{equation}
Adapting the technique of \cite{chalker} to our case of real-valued matrices, we find for such resolvents the general result
 \begin{equation}
 \label{reso1}
 \big\langle \text{Tr}\big[(z_1 - \bm{K})^{-1}(\bar{z}_2 - \bm{K}^{\rm T})^{-1}\big]\big\rangle = \frac{g_1 \bar{g}_2}{1-g_1 \bar{g}_2}
 \end{equation}
 where
 \begin{equation}
  g_1=\frac{z_1-\sqrt{z_1^2-4\eta}}{2\eta} \qquad \bar{g}_2=\frac{\bar{z}_2-\sqrt{\bar{z}_2^2-4\eta}}{2\eta}
 \end{equation}
Comparing with \eqref{eq:intresp}, one observes that $g_1$ and $\bar{g}_2$ are themselves response functions, with $z_1$ and $\bar{z}_2$ respectively replacing $z+\lambda$. In the case $\eta=0$, the r.h.s.\ of \eqref{reso1} simplifies further to $1/(z_1\bar{z}_2-1)$.

One expects the result \eqref{reso1} to apply whenever both $z_1$ and $\bar{z}_2$ are outside of the spectral ellipse. This is easily verified: one checks 
that $|g_1|=1$ is another parametrization for the boundary of this ellipse
 \begin{equation}
  |g_1|=1 \Leftrightarrow \quad \vline z_1-\sqrt{z_1^2-4\eta}\, \vline =2\eta
 \end{equation}
with foci $z_1=\pm 2\sqrt{\eta}$ and semi-axes $1+\eta$ and $1-\eta$ as before. So $z_1$ and $\bar{z}_2$ are outside of the spectral ellipse when $|g_1|<1$ and $|\bar{g}_2|<1$, which ensures that \eqref{reso1} is non-singular. The resolvent then diverges when e.g.\ $z_1=z_2$ and $z_1$ approaches the boundary of the ellipse.

 To work out the trace \eqref{eq:trace} defining the Laplace transformed correlation function, we need to set in the first resolvent \eqref{eq:res1} $z_1=\lambda-\text{i}\omega$ and $\bar{z}_2=\lambda+z$, and in the second \eqref{eq:res2} $z_1=\lambda-z$ and $\bar{z}_2=\lambda+\text{i}\omega$. After 
these substitutions the integration can be conveniently carried out using residues (see \ref{appendix:a}), with the result
    \begin{equation}
   \label{eq:fincor}
\tilde{C}(z)=\frac{\Sigma\,\tilde{R}(z)\tilde{R}(-z)}{1-\tilde{R}(z)\tilde{R}(-z)}
 \end{equation}
This is identical to the prediction  \eqref{eq:corple} of the Plefka approximation. Our conclusion is, therefore, that for our model with weak long-range interactions the extended Plefka approach provides fully exact results for response and correlation functions, in the thermodynamic limit $N\to\infty$.

\section{Quantitative results}

In this section we look at the quantitative results for our model system in more detail. As the model does not obey detailed balance when $\eta < 1$, we are in general 
dealing with a non-equilibrium steady state and will see some nontrivial features emerge from this.

We focus initially on the correlation function \eqref{eq:fincor}, which after substituting $\tilde{R}(z)$ and $\tilde{R}(-z)$ and simplifying reads
 \begin{equation}
 \label{FinCor}
  \tilde{C}(z)=\frac{4\,\Sigma}{\big[(\lambda+z)+\sqrt{(\lambda+z)^2-4\eta}\big]\big[(\lambda-z)+\sqrt{(\lambda-z)^2-4\eta}\big]-4}
 \end{equation}
Particular cases of note are
 \begin{equation}
\label{eq:CorAsym}
\tilde{C}(z)\big\rvert_{\eta=0}=
\frac{\Sigma}{\lambda^2-z^2-1}
 \end{equation}
 \begin{equation}
 \label{CorSym}
 \tilde{C}(z)\big\rvert_{\eta=1}=\Sigma\bigg[-\frac{1}{2} + \frac{1}{4 z}\sqrt{(\lambda+z)^2-4}-\frac{1}{4 z}\sqrt{(\lambda-z)^2-4}\bigg]
 \end{equation}
 \begin{equation}
 \tilde{C}(z)\big\rvert_{\eta=-1}=\Sigma\bigg[-\frac{1}{2} + \frac{1}{4 \lambda}\sqrt{(\lambda+z)^2+4}+\frac{1}{4 \lambda}\sqrt{(\lambda-z)^2+4}\bigg]
 \end{equation}
where the middle one is the detailed balance limit.

The long-time behaviour of $C(t-t')$ is determined by the singularities, i.e.\ poles and branch cuts, of $\tilde{C}(z)$ that are closest to the origin. It will be useful to think of 
these in relation to two copies of the spectral ellipse: bearing in mind that $\tilde{R}(\pm z)=g_1(z\mp\lambda)$, these are shifted to have their centres at $\pm \lambda$.

\begin{figure}
\includegraphics[width=\textwidth]{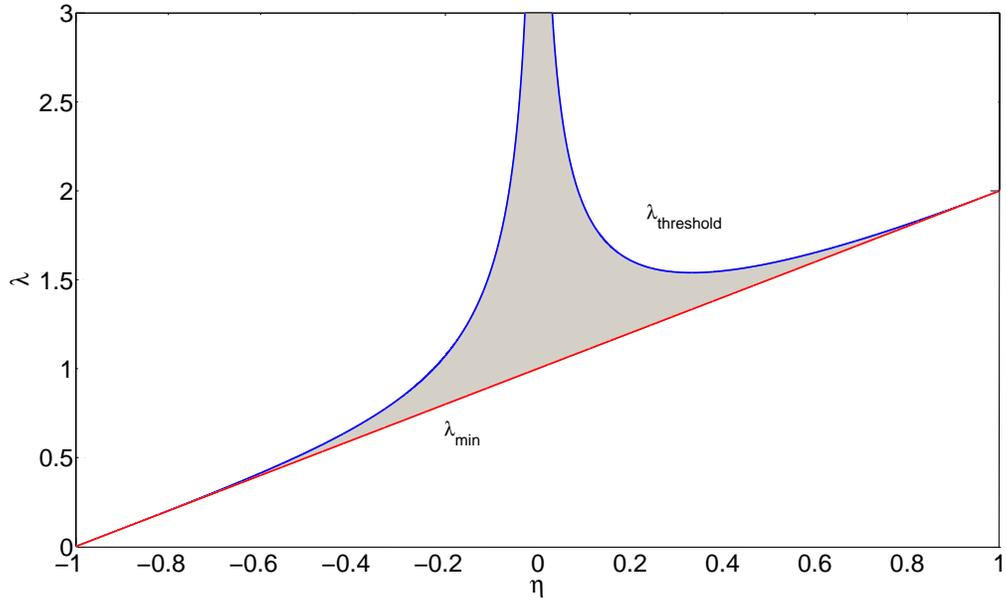}
\caption{$\lambda_{\text{threshold}}$ and $\lambda_{\rm min}$ as a function of $\eta$. The pole exists for $\lambda_{\rm min}<\lambda<\lambda_{\text{threshold}}$, i.e.\ for values of $\lambda$ lying in the grey shaded area. Where the pole exists it determines the asymptotic decay rate of the correlation function.}
\label{fig:l2th}
\end{figure}

For generic $\eta$ (see figures \ref{fig:symPo}, \ref{fig:symPoN} in \ref{appendix:a}), each of the two square roots in $\tilde{C}(z)$ contributes a branch cut. Each branch cut lies completely 
inside the relevant shifted spectral ellipse, and extends from one focus of the ellipse to the other. Explicitly, the branch cuts are
\begin{subequations}
\label{branchcuts}
\begin{align}
\pm\lambda-2\sqrt{\eta}<\,\text{Re}(z_{\text{bc}})<\pm\lambda+ 2\sqrt{\eta} \qquad \qquad \text{Im}(z_{\text{bc}})=0 \qquad &\eta>0\label{branchcuts_first}\\
\text{Re}(z_{\text{bc}}) =\pm\lambda   \qquad  -2\sqrt{|\eta|}<\text{Im}(z_{\text{bc}})<+ 2\sqrt{|\eta|} \qquad &\eta<0\label{branchcuts_second}
\end{align}
\end{subequations}
In the symmetric and anti-symmetric limits the ellipses degenerate to straight lines that coincide with the branch cuts while in the asymmetric case $\eta=0$ each branch cut shrinks to a point $z_{\text{bc}}=\pm \lambda$ at the centre of the spectral circle; see figure \ref{fig:AsyPo} in \ref{appendix:a}.
In addition to branch cuts, the Laplace transformed correlation function \eqref{FinCor} can have poles (for $\eta\neq 1,-1$). Setting the denominator of \eqref{FinCor} to zero gives
 \begin{equation}
 \label{zpole}
 z_{\text{pole}}=\pm z_0   \ \ {\rm with}\ \ z_0=  \bigg( \frac{1-\eta}{1+\eta} \bigg)\sqrt{\lambda^2-(1+\eta)^2} 
 \end{equation}
These poles emerge from the branch cuts as $\lambda$ is decreased below the threshold value $\lambda_{\text{threshold}}=(1+\eta)^2/(2\sqrt{\eta})$ for $\eta>0$ and $\lambda_{\text{threshold}}=(1-\eta^2)/(2\sqrt{|\eta|})$ for 
$\eta<0$ (see figure \ref{fig:l2th}); they do not exist for larger $\lambda$ because they are then no longer on the physical branch of $\tilde{C}(z)$. With decreasing $\lambda$ they then move towards the origin and reach it at a critical value for $\lambda$ given by $\lambda_{\text{min}}(\eta)=1+\eta$. 
This makes sense as the largest real part of eigenvalues within the spectral ellipse of $\bm{K}$ is exactly $1+\eta$: for $\lambda<\lambda_{\text{min}}$ these eigenvalues 
would cause the correlation function to diverge for long time differences.

The long-time or terminal decay rate $r$ of the correlation function is now given by the singularity, be it pole or branch cut edge, that has the smallest (positive) real part. Its inverse $1/r$ is the largest relaxation time.
The real part of the pole is $z_0$ itself, $r_{\text{pole}}=z_0$, while for the branch cut it is, from \eqref{branchcuts_first}, $r_{\text{bc}}=\lambda-2\sqrt{\eta}$ for $\eta>0$ and $r_{\text{bc}}=\lambda$ otherwise.
For $\lambda_{\text{min}}<\lambda<\lambda_{\text{threshold}}$, i.e.\ when the pole exists, one has $r_{\text{pole}}<r_{\text{bc}}$ thus $r_{\text{pole}}$ sets $r$. For all other values of $\lambda$, $r_{\text{bc}}=\lambda-2\sqrt{\eta}$ 
becomes responsible for the asymptotic decay. Bearing in mind that for a non-interacting system we would have $C(t-t')=\Sigma\exp(-\lambda |t-t'|)$, this means that the asymptotic decay rate of $C$ is only ever made smaller by the 
interactions, never larger.

\cleardoublepage

\subsection{Power Spectra and Power Laws}
\label{sec:PS}
We can obtain the power spectrum of the fluctuations in our system by setting $z=\text{i}\omega$ in \eqref{FinCor}, which converts the two-sided Laplace transform to a Fourier transform. For notational simplicity
we use the same symbol $\tilde{C}(\omega)$ for the latter as for the former, the meaning being clear from the argument of the function. Of primary interest is how the power spectrum differs from the simple Lorentzian
case corresponding to a purely exponential correlation function decay. 

We note first that the asymmetric case $\eta=0$ in \eqref{eq:CorAsym} always gives a Lorentzian power spectrum $\tilde{C}(\omega) = \Sigma/(\lambda^2-1+\omega^2)$. The presence of the interactions only manifests itself here in a change of the characteristic frequency from $\lambda$ to $r_{\text{pole}}=\sqrt{\lambda^2-1}$.
More generally for large $\lambda$ any non-trivial features of the correlation function will be hidden underneath a 
rapidly decaying $\exp(-\lambda |t-t'|)$ envelope, giving a Lorentzian power spectrum. This can be seen formally by taking $\lambda\to\infty$ in \eqref{FinCor} at $z$ of order $\lambda$.

\begin{figure}
\includegraphics[width=\textwidth]{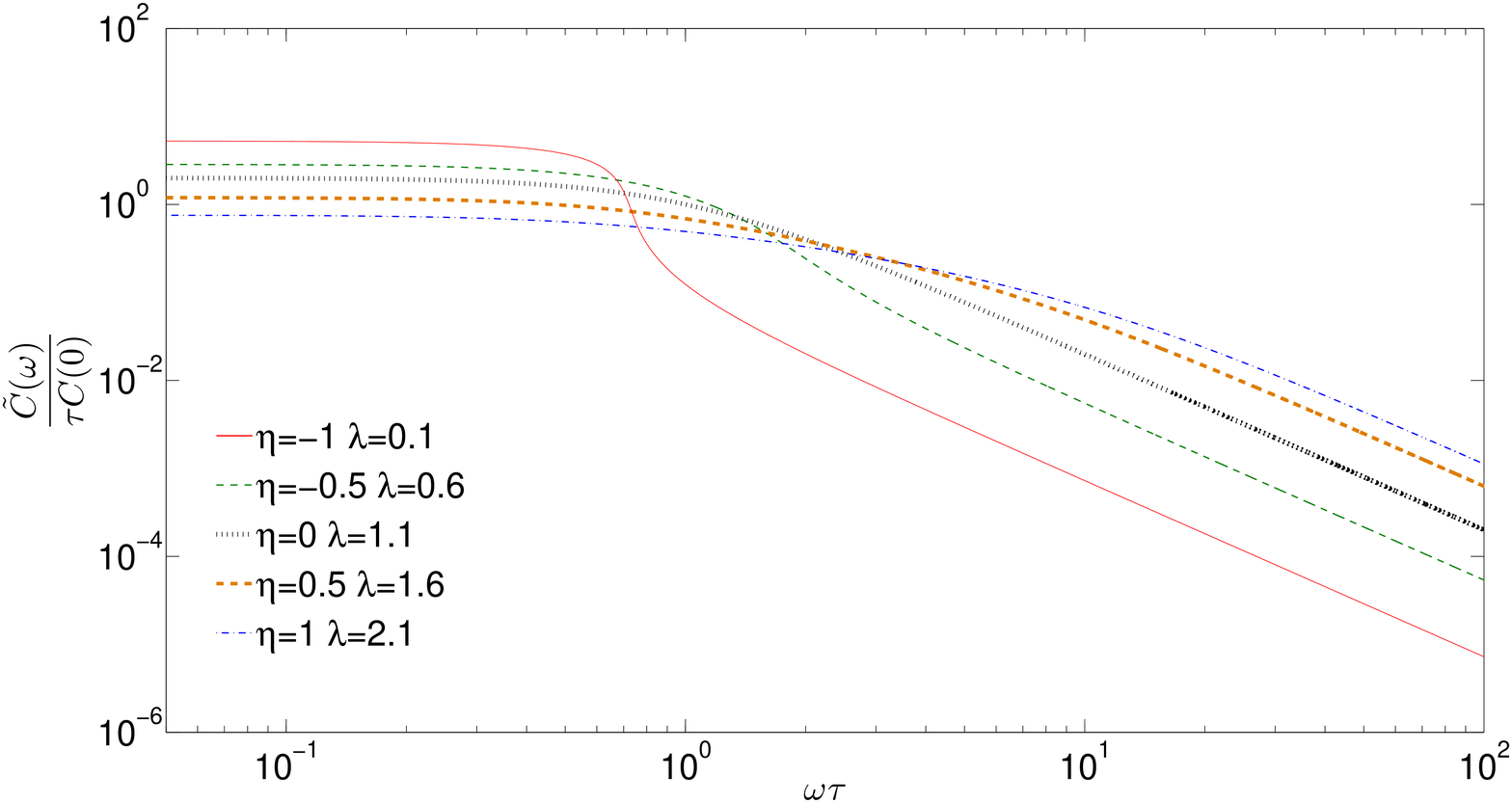}
\caption{Log-log plots of normalized (see text) power spectra for different symmetries. $\lambda$ is taken close to the corresponding minimal value $1+\eta$ to highlight non-Lorentzian features. 
For small $\omega$, the horizontal plateau represents an exponential cutoff, while the large $\omega$ tail $\sim 1/\omega^2$ is as for a Lorentzian ($\eta=0$). The power spectra for $\eta>0$ are broader than Lorentzian, suggesting slower decays that approach power laws for $\lambda\to 1+\eta$. For $\eta<0$ sharp drops in the power spectrum suggest oscillatory correlation decay in the time domain.}
\label{fig:Com5}
\end{figure}

Non-trivial power spectra are then expected to appear in the opposite regime of small $\lambda$, or more precisely small $\lambda - \lambda_{\text{min}}$ where $\lambda_{\text{min}}=1+\eta$. Keeping the self-interaction in the vicinity of this critical value allows one to detect interesting features such as power law tails, as illustrated in
figure \ref{fig:Com5}. To make the comparison of different spectral shapes easier it is convenient to remove uninteresting prefactors, i.e.\ to extract the overall scales of $\tilde{C}(\omega)$ and $\omega$ and plot the normalized quantities. For $\tilde{C}(\omega)$ we take as the scale
 \begin{equation}
  C(0)=\int_{-\infty}^{+\infty}\frac{d \omega}{2\pi}\,\tilde{C}(\omega)
 \end{equation}
 A scale for $\omega$ can be extracted as the inverse of a typical timescale $\tau$ for the decay of correlations; we choose in particular a root mean squared decay time 
 \begin{equation}
  \tau^2= \frac{\int_{-\infty}^{+\infty} dt \, t^2 C(t)}{2\int_{-\infty}^{+\infty} dt \, C(t)}=\frac{\int_{-\infty}^{+\infty} dt \, t^2 C(t)}{2\tilde{C}(0)}
 \end{equation}
 We then plot $\tilde{C}(\omega)/[\tau C(0)]$ versus $\omega \tau$ to ensure the normalized spectrum has a unit area under the curve. A log-log plot as in figure \ref{fig:Com5} shows clearly the large-frequency Lorentzian tail and suggests slower power law correlation decays for positive $\eta$ and oscillatory decay for negative $\eta$.
 
 \begin{figure}
\includegraphics[width=\textwidth]{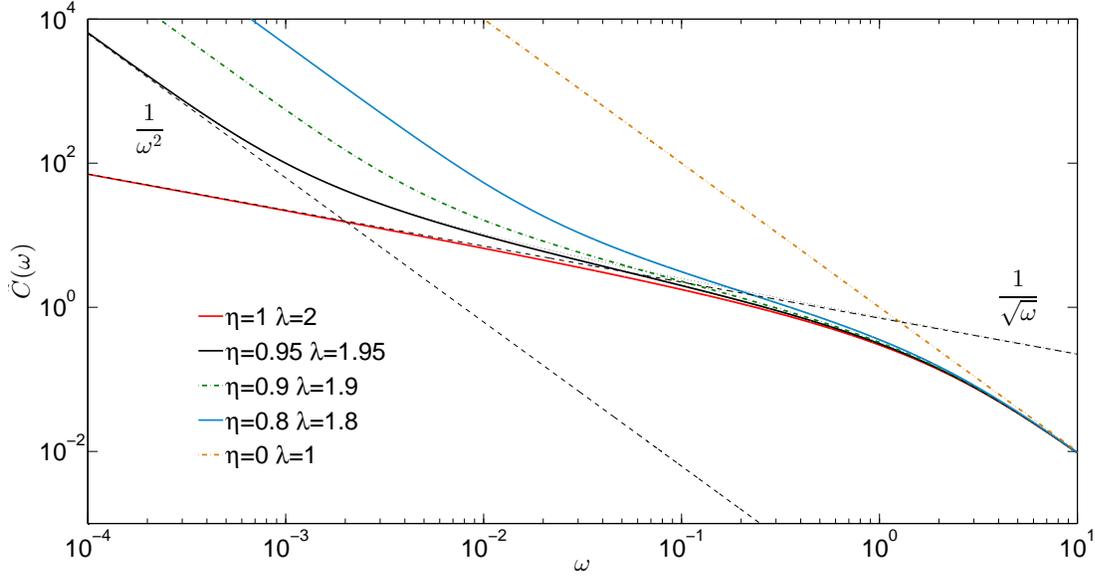}
\caption{Power spectrum at minimal $\lambda=1+\eta$ for positive symmetry parameters $\eta$. Dashed lines show the asymptotic power laws at small frequency, which govern the long-time behaviour. For slight asymmetry 
($\eta=1-\epsilon$), one sees interpolation between two master curves governing the frequency regimes of $\omega=O(1)$ and $\omega \sim \epsilon^2$. All curves show unnormalized power spectra, for noise amplitude $\Sigma=1$.}
 \label{fig:powerLawPaper01}
\end{figure}
\begin{figure}
\includegraphics[width=\textwidth]{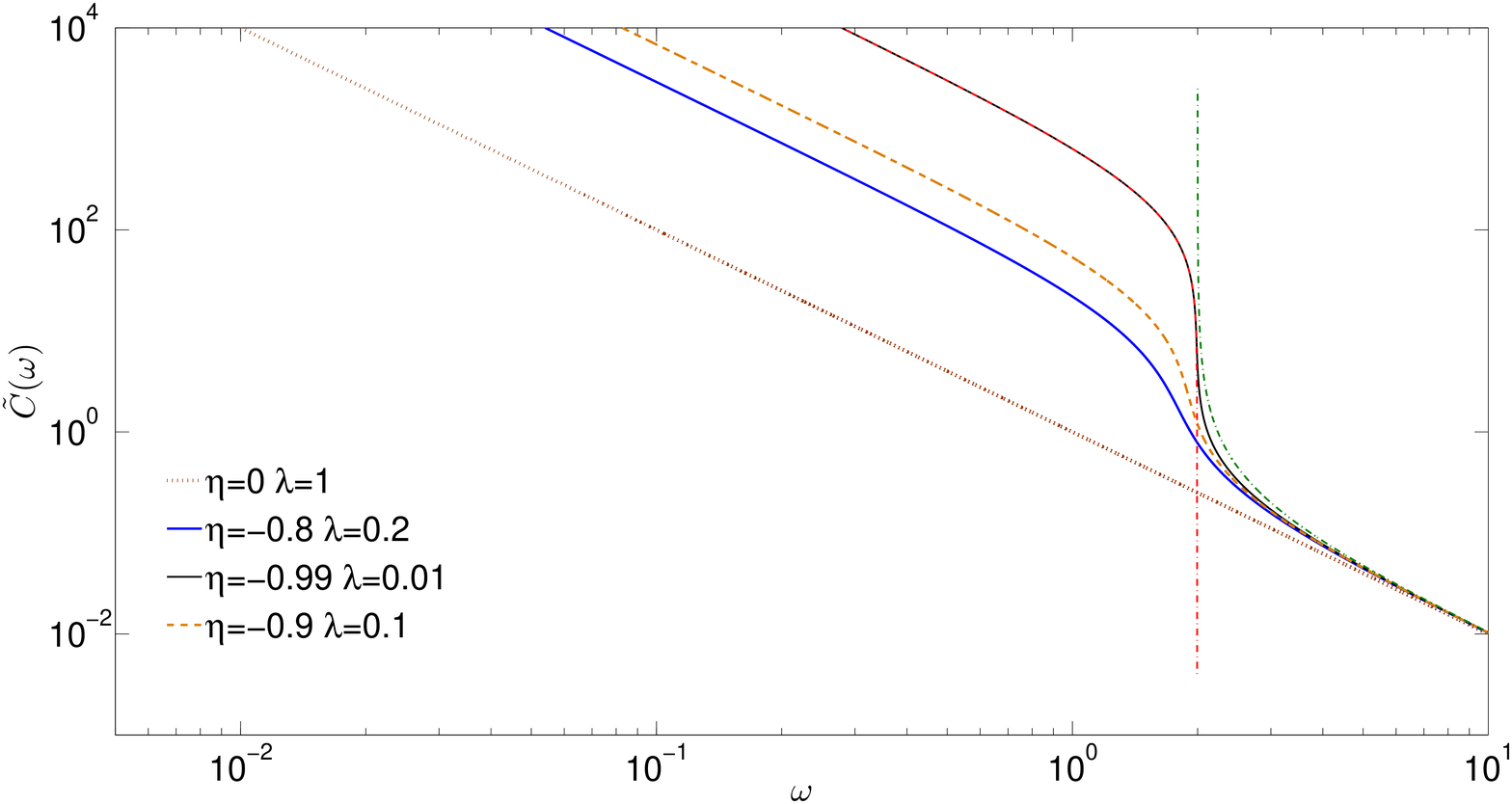}
\caption{Analogue of figure \ref{fig:powerLawPaper01} for negative symmetry parameters $\eta$. For small deviations from anti-symmetry ($\eta=-1+\epsilon$) the power spectrum splits into two regimes at $\omega=2$, each with its
own master curve (dashed lines). The amplitude in the low frequency part diverges as $1/\epsilon$ while the higher frequencies have a finite amplitude for $\epsilon\to 0$, so that an effective frequency cutoff at $\omega=2$ 
develops in the limit.}
 \label{fig:powerLawPaper02}
\end{figure}
We want to investigate more formally the emergence of power law behaviours for large time. This requires minimizing the effect of the exponential cut off provided by the self-interaction, so we consider $\lambda=\lambda_{\text{min}}$.
We then need to study the behaviour of $\tilde{C}(\omega)$ for small $\omega$. For $\eta=1$ one finds, by expansion of \eqref{CorSym}, $\tilde{C}(\omega)\sim 1/\sqrt{2\omega}$, corresponding to a $|t-t'|^{-1/2}$ decay in the time
domain. 
 
To understand the effect of slight deviations from symmetry we set $\eta=1-\epsilon$ with $\epsilon$ small. At fixed frequencies $\omega\sim O(1)$, the limit $\epsilon\to 0 $ then just 
retrieves $\tilde{C}(\omega)\vert_{\eta=1}$, so the latter
  \begin{equation}
\tilde{C}(\omega)\vert_{\eta=1}=-\frac{1}{2} + \frac{1}{2\sqrt{2}|\omega|}\sqrt{4-\lambda^2+\omega^2+\sqrt{\omega^4+2\omega^2(\lambda^2+4)+(\lambda^2-4)^2}}
 \end{equation}
evaluated at $\lambda=\lambda_{\text{min}}=2$ is the limiting ``master curve" for small $\epsilon$ in this part of the power spectrum
 \begin{equation}
\tilde{C}(\omega)\vert_{\eta=1,\,\lambda=2}=-\frac{1}{2} + \frac{1}{2\sqrt{2}|\omega|}\sqrt{\omega^2+\sqrt{\omega^4+16\omega^2}}
 \end{equation}
This master curve has asymptotic behaviour $\sim 1/\sqrt{2\omega}$ for small $\omega$, as found above: the power spectrum for small $\epsilon$ generically contains a non-Lorentzian power law regime as our initial numerics suggested.

If rather than fixing $\omega$ first and then taking $\epsilon\to 0$, we directly expand $\tilde{C}(\omega)$ for small $\omega$ at fixed $\epsilon$, we find
$\tilde{C}(\omega)\sim \frac{\epsilon^3}{2\omega^2}$ instead of $1/\sqrt{2\omega}$. Comparing the two expressions suggests that there is a crossover between two different regimes at a frequency scaling as $\epsilon^2$.  
To analyse the crossover region we therefore set $\omega=\epsilon^2 \gamma$ and take $\epsilon\to 0$ at fixed $\gamma$. The rescaled correlation $\epsilon \tilde{C}(\omega)$ then approaches a separate master curve 
 \begin{equation}
  \hat{C}(\gamma) = \frac{1+\sqrt{1+16\gamma^2}+ 2\sqrt{2}\gamma\sqrt{-1+\sqrt{1+16\gamma^2}}+\sqrt{2}\sqrt{1+\sqrt{1+16\gamma^2}}}{8\gamma^2\epsilon}
 \end{equation}
 The two tails of this low-frequency master curve retrieve the scalings found above as they should
\begin{subequations}
\begin{align}
&\gamma\ll1 \qquad (\omega\ll\epsilon^2) \qquad \hat{C}(\gamma)\sim 1/2\epsilon\gamma^2 \qquad \tilde{C}(\omega)\sim \epsilon^3/2\omega^2\\
&\gamma\gg1 \qquad (\omega\gg\epsilon^2) \qquad \hat{C}(\gamma)\sim 1/\epsilon\sqrt{2\gamma} \qquad \tilde{C}(\omega)\sim 1/\sqrt{2\omega}
\end{align}
 \end{subequations}
 The results of the above analysis are illustrated in figure \ref{fig:powerLawPaper01}. 
 Dashed lines indicate the exponents of the limiting power laws.

\newcommand{\lambdac}{\lambda_{\rm min}}

One notable aspect of the above power spectra is the $1/\omega^2$ tail for $\omega\to 0$, which makes the time-domain correlation function $C(t-t')$, obtained by inverse Fourier transform, formally infinite. 
This divergence could be regularized by taking $\lambda$ slightly larger than $\lambdac$; it turns out that in this limit the dominant contribution to $C(t-t')$ is from the pole $z_{\rm pole}$ defined 
in \eqref{zpole}. This contribution is of the order of $z_{\rm pole}^{-1} \exp[-z_{\rm pole}(t-t')]$, with $z_{\rm pole}$ scaling as $(\lambda-\lambdac)^{1/2}$.
 
Finally we consider the opposite end of the $\eta$ range and study the case of a slight deviation from antisymmetry, given by $\eta=-1+\epsilon$. To obtain the asymptotic behaviour for small $\epsilon$, we expand the power spectrum in $\epsilon$ and retain the two leading orders (which are $O(\epsilon^{-1})$ and $O(\epsilon^0)$). This yields
 \begin{equation}
\tilde{C}(\omega)\sim \frac{1}{\epsilon \omega^2}\bigg(4+2\sqrt{4-\omega^2}-\omega^2\bigg)+O(1) \qquad \omega<2
 \end{equation} 

 \begin{equation}
 \tilde{C}(\omega)\sim\frac{1}{\omega^2-4} +O(\epsilon) \qquad\omega>2
 \end{equation}
and these limiting curves are shown as dashed lines in figure \ref{fig:powerLawPaper02}. The key observation is that for small $\epsilon$ the power spectrum is confined almost entirely to the frequency range $0<\omega<2$, while higher frequencies are suppressed relative to this by a factor of $\epsilon$. As $\epsilon\to0$, a hard frequency cutoff therefore emerges at $\omega=2$.

\cleardoublepage
\subsection{Time Domain}
\label{sec:TD}
To gain further insight we can extract analytically the exact correlations in the time domain for $\eta=1$ (symmetric couplings) and $\eta=-1$ (anti-symmetric couplings).
For the symmetric case, using $\bm{K}^{\rm T}=\bm{K}$ in \eqref{eq:Cmatrix_generic_t_tt} gives
\begin{equation}
C(t,t') = 
\Sigma \int_0^{\text{min}(t,t')}dt'' \,\text{Tr}\,\text{e}^{(-\lambda+\bm{K})(t+t'-2t'')}
\end{equation}
The trace can be written as an integral over eigenvalues distributed according to Wigner's semi-circular law to give
\begin{eqnarray}
 C(t,t')&=\Sigma \int_{-2}^{2}\frac{dk}{2\pi}\sqrt{4-k^2}\,
\int_0^{\text{min}(t,t')}dt''\,\text{e}^{(-\lambda+k)(t+t'-2t'') }=\notag\\
 &= \Sigma\int_{0}^{\text{min}(t,t')}dt''\,\frac{I_1(2(t+t'-2t''))}{t+t'-2t''}\text{e}^{-\lambda(t+t'-2t'')}\\
 &=\Sigma\int_{|t-t'|}^{t+t'}dw\,\frac{I_1(2 w)}{2w}\text{e}^{-\lambda w} 
 \label{eq:mbf}
\end{eqnarray}
In the first step, we changed variable $k=2\cos{\theta}$ to write the $k$-integral as a modified Bessel function 
$\frac{I_1(\tau)}{\tau}=\frac{1}{\pi}\int_0^{\pi}d\theta(\sin{\theta})^2\text{e}^{\tau\cos{\theta}}$. The final equality follows by setting $w=t+t'-2t''$. In the long time limit the integral runs up to $+\infty$ and the result is manifestly TTI. The Fluctuation-Dissipation Theorem (FDT) \cite{FDT} is then expected to hold because for symmetric couplings the system has detailed balance. This can be checked by calculating the response function, which  comes out as simply the integrand of \eqref{eq:mbf}
\begin{equation}
 R(t-t')=\theta(t-t')\frac{I_1(2(t-t'))}{t-t'}\text{e}^{-\lambda(t-t')}
\end{equation}
This is as expected from the FDT $TR(t-t')=-(\partial/\partial t)C(t-t')$ where in our case $T=\Sigma/2$.

\begin{figure}
\includegraphics[width=\textwidth]{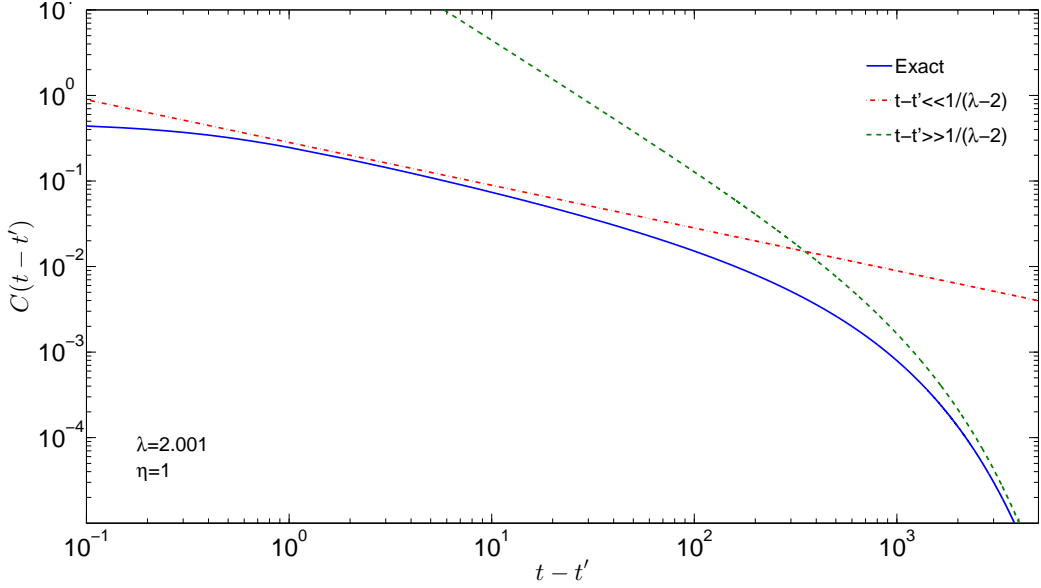}
\caption{Correlations in the time domain: comparison between the analytically exact expression and the asymptotic curves for small and large $t-t'$, for symmetric interactions, $\eta=1$.}
\label{fig:exasy01}
\end{figure}

The power law behaviour we found above in Fourier space corresponds to a power law in the time domain as can be confirmed using the asymptotic expression of the modified Bessel function
\begin{equation}
\label{eq:bessela}
 I_1(z)\sim \frac{\text{e}^z}{\sqrt{2\pi z}} \qquad z \gg 1
\end{equation}
As a consequence, the response decays asymptotically as
\begin{equation}
\label{eq:asyre}
 R(t-t')\sim \frac{\text{e}^{-(\lambda-2)(t-t')}}{\sqrt{4 \pi}(t-t')^{{3}/{2}}}
\end{equation}
For the correlation function, if we substitute the expression (\ref{eq:bessela}) into \eqref{eq:mbf} and carry out the integration we obtain the asymptotic behaviour
\begin{equation}
\label{eq:mbfasy}
 C(t-t')\sim \frac{1}{2\sqrt{\pi(t-t')}}F((\lambda-2)(t-t'))
\end{equation}
where 
\be
F(x)= \text{e}^{-x}-\sqrt{\pi x}\,\text{erfc}(x)
\ee
Two regimes can be distinguished: for $x\ll 1$ (i.e. $t-t'\ll 1/(\lambda-2)$) $F(x)\sim 1$ and one has $C(t-t')\sim 1/2\sqrt{\pi (t-t')}$, whereas
for $x\gg 1$ (i.e. $t-t'\gg 1/(\lambda-2)$) $F(x)\sim \text{e}^{-x}/2 x$ thus $C(t-t')\sim \text{e}^{- (\lambda-2)(t-t')}/4\sqrt{\pi}(t-t')^{{3}/{2}}(\lambda-2)$.
 A comparison between the exact \eqref{eq:mbf} and the asymptotic \eqref{eq:mbfasy} expressions for the correlation function is shown in figure \ref{fig:exasy01}.

\begin{figure}
\includegraphics[width=\textwidth]{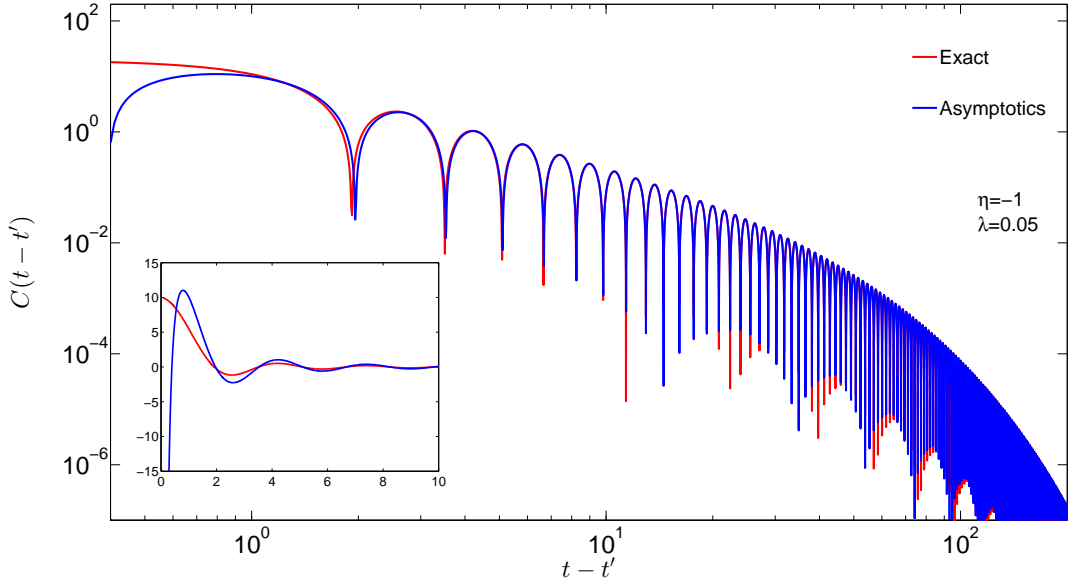}
\caption{Analogue of figure \ref{fig:exasy01} for antisymmetric interactions, $\eta=-1$. The power law decay is visible here in the envelope of the oscillatory relaxation.}
\label{fig:exasy02}
\end{figure}

If $\bm{K}$ is anti-symmetric ($\eta=-1$), one can perform largely analogous calculations. The explicit expression for the correlations is
\begin{equation}
C(t,t') = \Sigma \int_0^{\text{min}(t,t')}dt'' \,\text{Tr}\,\text{e}^{-\lambda(t+t'-2t'')+\bm{K}(t-t')}=\frac{\Sigma}{2\lambda}(\text{e}^{-\lambda|t-t'|}-\text{e}^{-\lambda(t+t')})\text{Tr}\,\text{e}^{\bm{K}(t-t')}
\end{equation}
Replacing the trace by an integral over the eigenvalue spectrum, which is now a Wigner semicircle rotated onto the imaginary axis, and taking the long-time limit gives the TTI form
\begin{equation}
\label{eq:mbf1}
C(t-t')= \frac{\Sigma}{2\lambda}\frac{J_1(2(t-t'))}{t-t'}\text{e}^{-\lambda(t-t')}
\end{equation}
The Bessel function of the first kind in this is related to the modified Bessel function by $J_1(\text{i}x)=i I_1(x)$ \cite{abramowitz}. 
The response function for $t>t'$ is found similarly as
\begin{equation}
\label{eq:resasym}
 R(t-t')=\theta(t-t')\frac{J_1(2(t-t'))}{t-t'}\text{e}^{-\lambda(t-t')}
\end{equation}
From the asymptotics of $J_1$ one then finds for large time differences
\begin{equation}
\label{eq:mbf1asy}
 C(t-t')\sim \frac{\text{e}^{-\lambda(t-t')}\,\sin{\big[2\big((t-t') -\frac{\pi}{8}\big)\big]}}{\lambda \sqrt{\pi}(t-t')^{{3}/{2}}}
\end{equation}
so the power law component of the decay is as for the symmetric case $\eta=1$, but here with an oscillatory modulation from the exponential.
For a comparison between the exact \eqref{eq:mbf1} and the asymptotic \eqref{eq:mbf1asy} expressions for the correlation we refer to figure \ref{fig:exasy02}.

The results \eqref{eq:mbf1} and \eqref{eq:resasym} show that correlation and response are fully proportional for $\eta=-1$. This is unexpected from the point of view of the FDT, but of course here we are considering interactions that are not symmetric. Probability currents are then generically present in the steady state. These translate into  additional terms in the FDT, giving rise to a Modified Fluctuation-Dissipation Theorem; see e.g.~\cite{chetrite}. One can check that these terms generate exactly the proportionality between correlation and response we found above (see \cite{mythesis} for details).
 
\section{Discussion and Conclusion}
\label{sec:conclusions}
In this paper we have developed and studied a novel approach for deriving approximate descriptions for large dynamical systems with continuous degrees of freedom. We refer to the method as an ``extended Plefka expansion", where the
extension lies in including second order statistics of the fluctuating degrees of freedom in the set of order parameters, rather than only first order averages, i.e.\ means.
Expanding in second order of interaction strength, we derive from the original dynamics -- a system of coupled stochastic differential equations -- effective equations of motion for each
single degree of freedom. These equations are decoupled, with interactions being represented by effective noise that is no longer white, and a memory term that connects each degree of 
freedom to its own past. The parameters governing these effective interaction terms are obtained from deterministic (nonlinear) coupled equations.

One key question we studied is under what circumstances the extended Plefka expansion can give exact results for large systems. We demonstrated explicitly for a linear dynamical model that this 
exactness holds when couplings are of mean field type, i.e.\ weak and long-ranged. An analogy can be drawn with works on soft spins dynamics \cite{sompolinsky0} \cite{sompolinsky1}, where the exact 
infinite-range limit produces local mean field equations with self-consistent propagator and noise.
Importantly, the agreement we show holds independently of whether the dynamics obeys detailed balance, due to symmetry in the interaction 
coefficients, or not; we explored the entire range of symmetry parameters from symmetry ($\eta=1$) to asymmetry ($\eta=0$) to anti-symmetry ($\eta=-1$). We also studied the quantitative features of the model in some detail, focussing on correlation functions and power spectra as their Fourier transform; this analysis revealed non-trivial crossover phenomena in the vicinity of either full symmetry or full anti-symmetry.

The extended Plefka method makes exact predictions for our linear model system, whereas -- as we discussed -- a conventional Plefka expansion fails to predict any non-trivial effects in correlations and 
responses. This suggests our method as a promising candidate for the accurate reconstruction of the dynamics of large systems also in generic nonlinear settings that cannot be solved analytically. 
The equations we have derived can be applied directly to such a generic case. We have mostly restricted ourselves to a model without self-interactions beyond the basic linear one that we assume, 
but this restriction can easily be lifted at the expense of longer expressions for the memory functions and effective noise correlations (see \ref{sec:complete_TAP} for a summary and \cite{mythesis} for details). 

An important question for such future applications is in what other scenarios one would expect the extended Plefka method to become exact in the large system limit. 
Generalizing from our linear model one could consider e.g.\ nonlinear drift terms of the form $\phi_i(\bm{x})= \sum_j K_{ij}g(x_j)$, where $g(\bm{x})$ is a generic non-linear function. With the potential application in biochemical networks in mind, $K_{ij}g(x_j)$ could describe the interaction due to reactions between different species $i$ and $j$, $K_{ij}$ being a reaction rates. From central limit theorem arguments one would expect that the dynamics of such a nonlinear system would again be described exactly by the extended Plefka method, provided that the $K_{ij}$ are weak and long-ranged. This should hold even if the nonlinearities are made species-specific so that $g(x_j)$ is replaced by $g_j(x_j)$.
Related models can be found  in the context of neural networks,  where $g(x_j)$ plays the role of a nonlinear gain function combining ``inputs'' to determine certain ``outputs". 
The mean field properties of such models, in the case of asymmetric $K_{ij}$ and $g(x_j)$ of sigmoid shape, were studied by Sompolinsky and coworkers \cite{sompolinsky2} and are consistent 
with the extended Plefka predictions. In general one could think of other simple scenarios where some moments of the variables $x_i$ can be calculated exactly and these may also provide useful 
future testbeds for our method. Interestingly, after the completion of this work, we discovered that an alternative perturbative approach also taking into account second 
moments had already been applied  by Biroli in the derivation of dynamical TAP equations for the $p$-spin spherical model~\cite{biroli}.
These TAP equations are the fixed-disorder analogue of the disorder-averaged equations first derived by \cite{thirumalai}. 
We have checked that the extended Plefka expansion gives back exactly Biroli's equations when applied to the $p$-spin model, with the Lagrange multiplier for the spherical constraint playing the role of our $\lambda_i$;
see \ref{sec:complete_TAP}. This is an important consistency check. Nevertheless we stress that the framework discussed in this paper is in principle wider, encompassing generic continuous variables and generic nonlinear
interaction terms. In addition, it is aimed at producing approximate decoupled equations that could be regarded as the first step for implementing inference algorithms.

A promising further development of our method would be to find a more sophisticated treatment of nonlinear self-interactions. In our present approach, these would be subsumed into the general interaction terms. Alternatively one could try to treat nonlinear self-interactions exactly, by keeping them as part of the non-interacting baseline for the Plefka expansion. This would result in effective equations of motion that are still decoupled but now nonlinear and driven by memory terms and coloured noise. The resulting self-consistency conditions for the order parameters would then have to be obtained by simulation, but there are precedents \cite{manfred} for doing this in a computationally efficient manner.

A further direction for future work would be to understand in more detail the relation to the Expectation-Propagation (EP) algorithm \cite{opperwinther}. For the case of linear self-interactions $-\lambda_i x_i$ that we mostly focussed on, EP and our extended Plefka method both yield factorized (over degrees of freedom) probability distributions over system trajectories, with the same non-interacting Gaussian baseline. It would therefore be interesting to clarify what the differences between the approaches are and under what circumstances they might lead to identical approximations. 

One important simplification we had to make was to assume that the different degrees of freedom $x_i$ are affected by independent noise, so that the noise covariance matrix 
$\bm{\Sigma}$ is diagonal. On the other hand,  in biochemical networks there are generically off-diagonal noise correlations: noise arises from the stochasticity of when reactions take place, and each non-trivial reaction affects the number of molecules from several molecular species.
The extension of our approach to this case requires further work. If the noise covariance $\bm{\Sigma}$ is at least independent of the state $\bm{x}$ of the system -- though even this is not the generic case for reaction networks -- then one could imagine transforming the variables $x_i$ linearly to diagonalize $\bm{\Sigma}$. This would then make our approach directly applicable, but would also make the biological interpretation of any predictions rather less intuitive.

In the long term our approximation framework should also help one to tackle network reconstruction problems, and this is a further important direction for future work. 
In fact, once the forward dynamics has been fully characterized as we have done here, one can think of setting up inverse techniques based on the same description. This would allow one e.g.\ to infer 
the states of hidden (unobserved) nodes from observations of other (visible) variables \cite{kalman}, and ultimately to learn interaction parameters and hence network structure from data.

\ack
This work is supported by the Marie Curie Training Network NETADIS (FP7, grant 290038). We thank Ludovica Bachschmid Romano, Zdzis\l{}aw Burda and Yasser Roudi for insightful discussions.

\cleardoublepage
\appendix
\section{Residue calculation}
\label{appendix:a}
\begin{figure}
 \begin{center}
 \begin{tikzpicture} 
\draw [help lines,->] (0,-3) -- (0,3);  
\draw [help lines,->] (-5,0) -- (5,0);   
\node at (5,-0.3){$\text{Re}(z)$};
\node at (-0.6,2.8) {$\text{Im}(z)$};
  \node at (3,0) {$\bm{|}$};
  \node at (-3,0) {$\bm{|}$};
\node at (2.3,0) {$\times$};
  \node at (-2.3,0) {$\times$};
  \node at (3.1,-0.4){\small{$\lambda$}};
  \node at (2.2,-0.4){\small{$\sqrt{\lambda^2-1}$}};
  \node at (-1.9,-0.4){\small{$-\sqrt{\lambda^2-1}$}};
  \node at (-3.1,-0.4){\small{$-\lambda$}};
 \draw[style=dashed] (3,0) circle (1.5);
 \draw[style=dashed] (-3,0) circle (1.5);
\end{tikzpicture}
  \caption{Singularities of $\tilde{C}(z)$ in the complex $z$-plane for asymmetric interactions ($\eta=0$): for this value of $\eta$ the only singularities are the two poles $z_{\text{pole}}=\pm\sqrt{\lambda^2-1}$. The random matrix calculation following \cite{chalker}, which uses a perturbative approach, applies only outside the two copies of the spectral circle (of radius one) shifted to be centred at $z=\pm\lambda$, but the results can be continued analytically into the circles as $\tilde{C}(z)$ is a Laplace transform.}
\label{fig:AsyPo}
\end{center}
\end{figure}
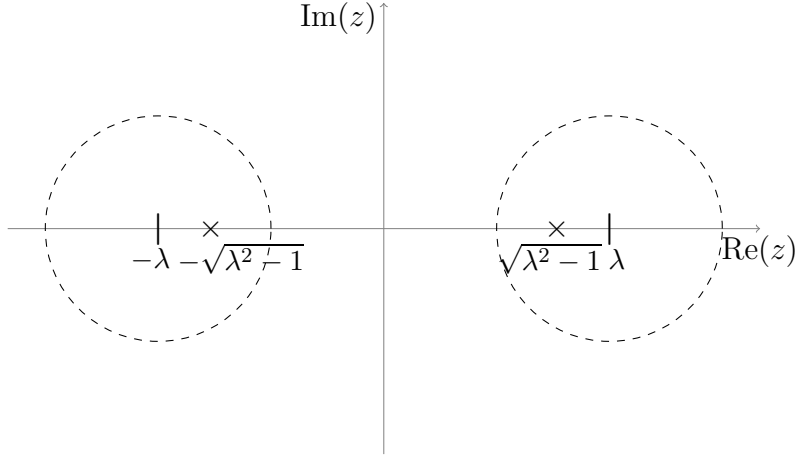

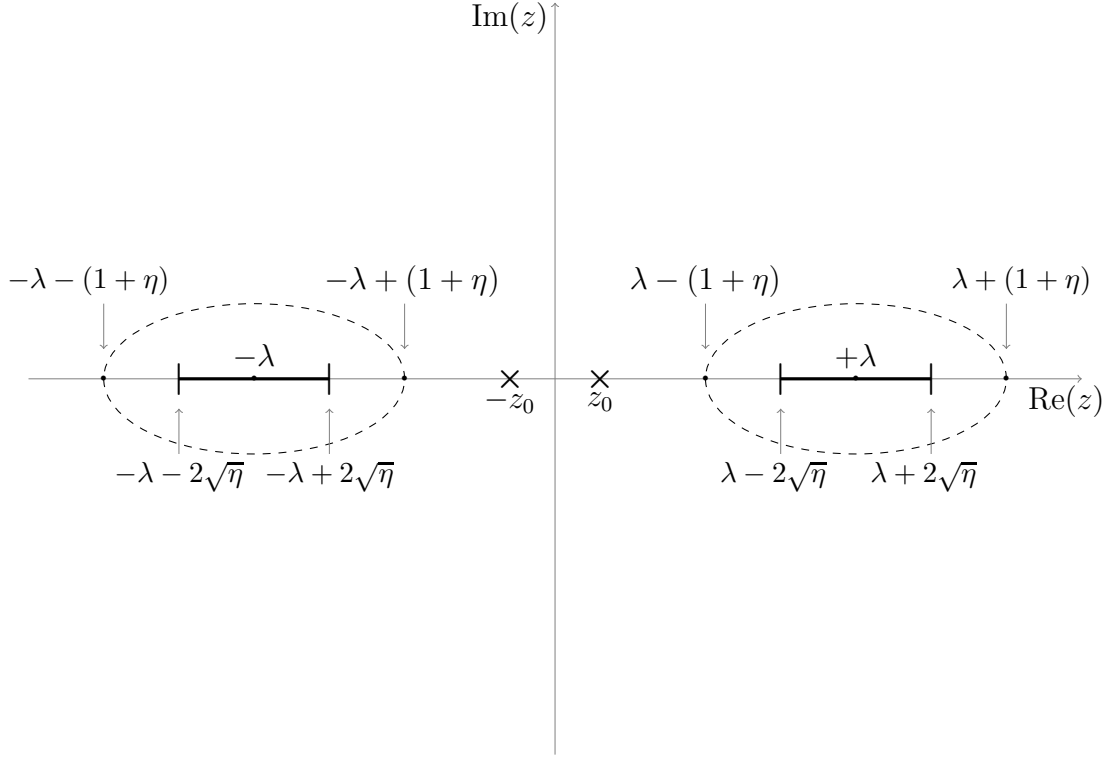
\begin{figure}
 \begin{center}
 \begin{tikzpicture}
  \draw[style=dashed] (4,0) ellipse (2 and 1);
 \draw[style=dashed] (-4,0) ellipse (2 and 1);
\draw [help lines,->] (0,-5) -- (0,5);  
\draw [help lines,->] (-7,0) -- (7,0);   
\node at (6.8,-0.3){$ \text{Re}(z)$};
\node at (-0.6,4.8) {$ \text{Im}(z)$};
\node at (-0.6,0) {$\bm{\times}$}; 
\node at (0.6,0) {$\bm{\times}$};
\node at (0.6,-0.3){$z_0$};
\node at (-0.6,-0.3){$-z_0$};
 \draw[very thick,black]
  (-3,0) -- (-5,0); 
  \draw[very thick,black]
  (3,0) -- (5,0);

  \node at (-4,0) {$\bm{\cdot}$};
  \node at (4,0) {$\bm{\cdot}$};
   \node at (-4,0.3) {$-\lambda$};
  \node at (4,0.3) {$+\lambda$};
    \node at (6,0) {$\bm{\cdot}$};
    \node at (-6,0) {$\bm{\cdot}$};
    \node at (2,0) {$\bm{\cdot}$};
    \node at (-2,0) {$\bm{\cdot}$};
     \node at (2,1.3) {$\lambda-(1+\eta)$};
     \draw [help lines,->] (2,1.0) -- (2,0.4); 
    \node at (-1.9,1.3) {$-\lambda+(1+\eta)$};
     \draw [help lines,->] (-2,1.0) -- (-2,0.4);
  \node at (5,0) {$\bm{|}$};
    \node at (3,0) {$\bm{|}$};
      \node at (-5,0) {$\bm{|}$};
        \node at (-3,0) {$\bm{|}$};
    \node at (6.2,1.3){\small{$\lambda+ (1 +\eta)$}}; 
    \draw [help lines,->] (6,1.0) -- (6,0.4); 
    \node at (-6.2,1.3){\small{$-\lambda- (1 +\eta)$}};
    \draw [help lines,->] (-6,1.0) -- (-6,0.4);
  \node at (4.9,-1.3){\small{$\lambda+2 \sqrt{\eta}$}};
  \draw [help lines,->] (5,-1.0) -- (5,-0.4);
  \node at (2.9,-1.3){\small{$\lambda-2 \sqrt{\eta}$}};
  \draw [help lines,->] (3,-1.0) -- (3,-0.4); 
  \node at (-3,-1.3){\small{$-\lambda+2 \sqrt{\eta}$}};
  \draw [help lines,->] (-3,-1.0) -- (-3,-0.4);
  \node at (-5,-1.3){\small{$-\lambda-2 \sqrt{\eta}$}};
  \draw [help lines,->] (-5,-1.0) -- (-5,-0.4);
\end{tikzpicture}
\caption{Singularities of $\tilde{C}(z)$ in the complex $z$-plane for generic positive interaction symmetry ($\eta>0$): there are two  poles $z_{\text{pole}}=\pm z_0$ as well as two branch cuts connecting the four points $z_{\text{bc}}=\pm \lambda\pm 2\sqrt{\eta}$. The random matrix calculation applies only outside the two copies of the spectral ellipse shifted to be centred at $z=\pm\lambda$.
The ellipses have real and imaginary semi-axes $1+\eta$ and $1-\eta$, respectively; the foci are the edges of the branch cuts.}
\label{fig:symPo}
\end{center}
\end{figure}

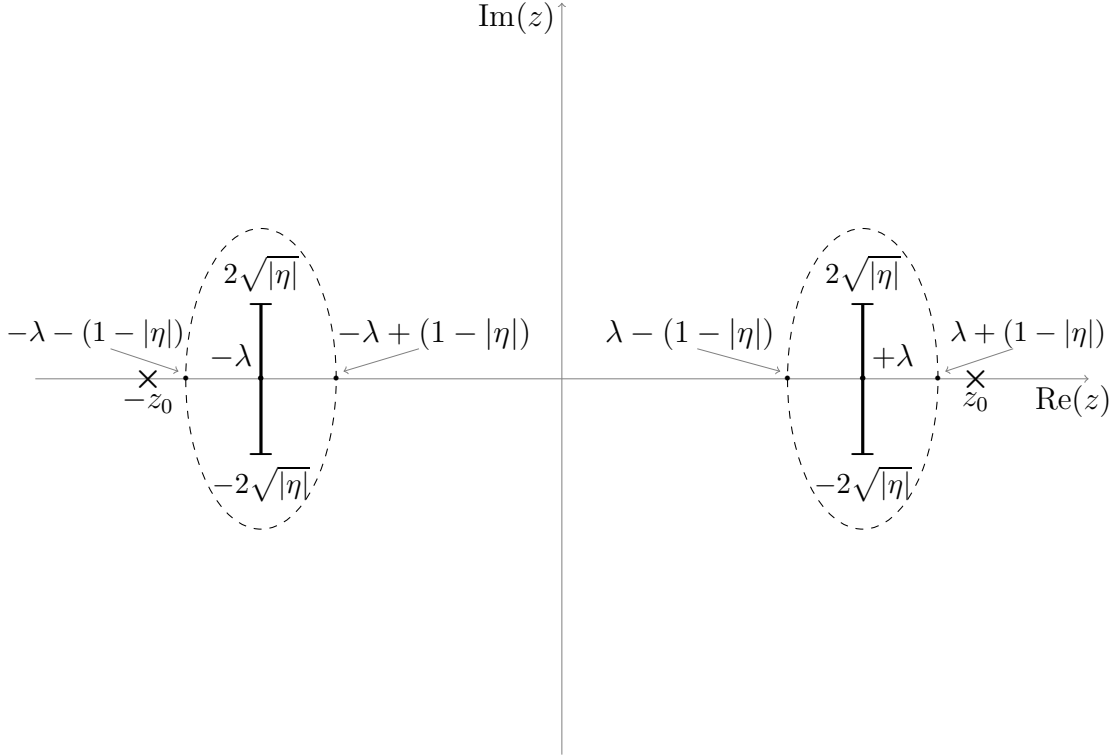
\begin{figure}
 \begin{center}
 \begin{tikzpicture}
  \draw[style=dashed] (4,0) ellipse (1 and 2);
  \draw[style=dashed] (-4,0) ellipse (1 and 2);
\draw [help lines,->] (0,-5) -- (0,5);  
\draw [help lines,->] (-7,0) -- (7,0);   
\node at (6.8,-0.3){$ \text{Re}(z)$};
\node at (-0.6,4.8) {$ \text{Im}(z)$};
\node at (-5.5,0) {$\bm{\times}$}; 
\node at (5.5,0) {$\bm{\times}$};
\node at (5.5,-0.3){$z_0$};
\node at (-5.5,-0.3){$-z_0$};
 \draw[very thick,black]
  (-4,-1) -- (-4,1); 
  \draw[very thick,black]
  (4,-1) -- (4,1);
  \node at (-4,0) {$\bm{\cdot}$};
  \node at (4,0) {$\bm{\cdot}$};
   \node at (-4.4,0.3) {$-\lambda$};
  \node at (4.4,0.3) {$+\lambda$};
    \node at (4,0) {$\bm{\cdot}$};
    \node at (-4,0) {$\bm{\cdot}$};
    \node at (3,0) {$\bm{\cdot}$};
    \node at (-5,0) {$\bm{\cdot}$};
    \node at (5,0) {$\bm{\cdot}$};
    \node at (-3,0) {$\bm{\cdot}$};
     \node at (1.7,0.6) {$\lambda-(1-|\eta|)$};
      \draw [help lines,->] (1.8,0.4) -- (2.9,0.1); 
    \node at (-1.7,0.6) {$-\lambda+(1-|\eta|)$};   
     \draw [help lines,->] (-1.9,0.4) -- (-2.9,0.1); 
  \node at (-4,1) {$\bm{-}$};
    \node at (-4,-1) {$\bm{-}$};
      \node at (4,1) {$\bm{-}$};
        \node at (4,-1) {$\bm{-}$};
    \node at (6.2,0.6){\small{$\lambda+ (1 -|\eta|)$}};  
     \draw [help lines,->] (6.,0.4) -- (5.1,0.1); 
    \node at (-6.2,0.6){\small{$-\lambda- (1 -|\eta|)$}};  
    \draw [help lines,->] (-6.,0.4) -- (-5.1,0.1); 
  \node at (-4,-1.4){\small{$-2\sqrt{|\eta|}$}};
  \node at (-4,1.4){\small{$2 \sqrt{|\eta|}$}};
 \node at (4,-1.4){\small{$-2\sqrt{|\eta|}$}};
  \node at (4,1.4){\small{$2 \sqrt{|\eta|}$}};
\end{tikzpicture}
  \caption{Analogue of figure \ref{fig:symPo} for negative symmetry parameters $\eta$. The branch cuts and major semi-axes of the ellipses are now along the imaginary rather than the real direction, so that the foci and hence the edges of the branch cuts are at $z=\pm\lambda\pm2\text{i}\sqrt{|\eta|}$.}
\label{fig:symPoN}
\end{center}
\end{figure}

\setcounter{section}{1}
We provide some details here of the exact calculation of the Laplace transformed correlation function $\tilde{C}(z)$ for the linear model with weak long-range interactions. The singularity structure of this function in the complex $z$-plane is sketched in figures \ref{fig:AsyPo}, \ref{fig:symPo} and \ref{fig:symPoN}. Below it will be useful to remember also that the singularities of the response function $\tilde{R}(z)$ are the same as those singularities of $\tilde{C}(z)$ that lie in the left half-plane. The difference arises because in the time domain the correlation function $C(t-t')$ is even in $t-t'$, while the response $R(t-t')$ vanishes for $t-t'<0$. 

Let us begin with asymmetric random interactions ($\eta=0$), for which \eqref{firMan} can be decomposed as
   \begin{equation}
  \text{Tr}\,\tilde{\bm{C}}(z) =\Sigma (I_1+I_2)
 \end{equation}
 where
 \begin{equation}
  I_1 = \int_{-\infty}^{+\infty}\frac{d \omega}{2\pi}\,\frac{1}{z-\text{i}\omega}\frac{1}{1-(\lambda+z)(\lambda-\text{i}\omega)} 
 \end{equation}
 \begin{equation}
  I_2 = \int_{-\infty}^{+\infty}\frac{d \omega}{2\pi}\,\frac{1}{z-\text{i}\omega}\frac{1}{(\lambda-z)(\lambda+\text{i}\omega)-1} 
 \end{equation}
 
 \begin{figure}[h]
 \begin{center}
 \begin{tikzpicture}
\draw [help lines,->] (0,-5) -- (0,5);  
\draw [help lines,->] (-5,0) -- (5,0);   
\node at (4.8,-0.3){$\text{Re}(\omega)$};
\node at (-0.6,4.8) {$\text{Im}(\omega)$};
\node at (-2,4.3) {$\mathcal{C}$};
\node at (0,-4) {$\bm{\times}$}; 
\node at (0,2.5) {$\bm{\times}$};
\node at (0,-1.9) {$\bm{\times}$};
\node at (0.5,-4){$\omega_1$};
\node at (0.5,-1.9){$\omega_2$};
\node at (0.5,2.5){$\omega_3$};
\node at (-4.7,4.7){$I_1+I_2$};
\draw[very thick,black,xshift=2pt,
decoration={ markings,  
      mark=at position 0.2 with {\arrow{latex}}, 
      mark=at position 0.6 with {\arrow{latex}},
      mark=at position 0.8 with {\arrow{latex}}, 
      mark=at position 0.98 with {\arrow{latex}}}, 
      postaction={decorate}]
  (-4.5,0) -- (4.5,0);
\draw[thick,black,xshift=2pt,
decoration={ markings,
      mark=at position 0.2 with {\arrow{latex}}, 
      mark=at position 0.4 with {\arrow{latex}},
      mark=at position 0.6 with {\arrow{latex}}, 
      mark=at position 0.8 with {\arrow{latex}}}, 
      postaction={decorate}]
 (4.5,0) arc (0:180:4.5) ;

\end{tikzpicture}
  \caption{Integration contour in the complex plane for $I_1+I_2$ in the asymmetric case.}
\label{fig:Iasy}
\end{center}
\end{figure}
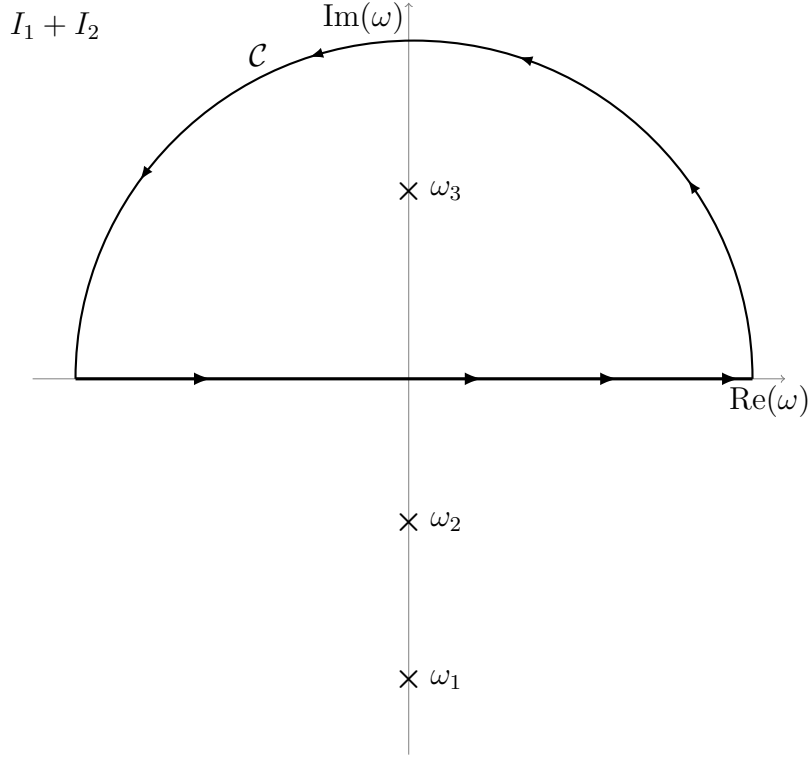
 These integrals can be performed in the complex plane as parts of integrals along a closed path. In fact, if we denote the integrands as $f_{1,2}(\omega)$, we can write
   \begin{equation}
I_{1,2}=\int_{-\infty}^{+\infty} \frac{d \omega}{2\pi} f_{1,2}(\omega)= \big( \int_{-\infty}^{+\infty}+\int_{\mathcal{C}} \big) \frac{d \omega}{2\pi} f_{1,2}(\omega)= \oint \frac{d \omega}{2\pi} f_{1,2}(\omega) = \frac{1}{2\pi} \,2\pi \text{i} \sum_i \text{Res} f_{1,2}(\omega)\vert_{\omega=\omega_i} 
 \end{equation}
Here the $\omega_i$ refer to the poles inside the closed path, as drawn in figure \ref{fig:Iasy}. The value of the integral along the semicircle $\mathcal{C}$ vanishes when the radius goes to infinity as $f_{1,2}(\omega)\approx\frac{1}{\omega^2}\rightarrow 0$ for $|\omega|\rightarrow\infty$. 
The poles for $f_1(\omega)$ are
\begin{equation}
 \omega_1=-\text{i}z \qquad \omega_2= \text{i}\,\bigg[\frac{1-\lambda(\lambda+z)}{\lambda+z}\bigg]
\end{equation}
while the poles for $f_2(\omega)$ are
\begin{equation}
 \omega_1=-\text{i}z \qquad \omega_3= -\text{i}\,\bigg[\frac{1-\lambda(\lambda-z)}{\lambda-z}\bigg] 
\end{equation}
To locate the poles in the complex $\omega$-plane we can fix a convenient region for the value of $z$, which for the purposes of our integration is an external parameter, and then continue the result analytically in $z$ at the end. In particular, it is useful to ensure that $z+\lambda$ and $\lambda-z$ are kept outside the support of the spectrum of the interaction matrix $\bm{K}$, i.e.\ that $z$ stays outside the circles in figure \ref{fig:AsyPo}.
Let us therefore choose $z$ as real and $z>\lambda+1$. With this choice, $\omega_1$ and $\omega_2$ lie on the negative imaginary axis, while $\omega_3$ lies on the positive axis. We thus close the integration contour in the upper half plane so that 
it includes only $\omega_3$ and obtain
  \begin{equation}
I_1+I_2 = 2\pi \text{i} \,\text{Res} f_1(\omega)\vert_{\omega=\omega_2}= \frac{1}{2\pi} \,2\pi \text{i} \,  \frac{1}{\text{i}\,(\lambda^2-z^2-1)} =\frac{1}{\lambda^2-z^2-1}
 \end{equation}
 Thus
 \begin{equation}
 \label{CorAsym}
  \tilde{C}(z)=\frac{\Sigma}{\lambda^2-z^2-1}
 \end{equation}
This exact result agrees with the prediction \eqref{eq:corple} of the extended Plefka method once we insert the appropriate expression \eqref{reAsPl} for the response in the asymmetric case.

For arbitrary correlations $\eta$ between $K_{ij}$ and $K_{ij}$, \eqref{eq:trace} can be rewritten as
\begin{eqnarray}
   \label{eq:gen}
\fl\tilde{C}(z)&=\Sigma\int_{-\infty}^{+\infty} \frac{d \omega}{2\pi} \big[\tilde{R}(-z)f_1(\omega)+\tilde{R}(z)f_2(\omega)\big]= \notag\\
\fl&=\Sigma\bigg(-\oint_1 \frac{d \omega}{2\pi} \tilde{R}(-z)f_1(\omega)+\oint_2 \frac{d \omega}{2\pi}\tilde{R}(z)f_2(\omega)+\int_{\mathcal{C}_1} \frac{d \omega}{2\pi}\tilde{R}(-z)f_1(\omega)-\int_{\mathcal{C}_2} \frac{d \omega}{2\pi}\tilde{R}(z)f_2(\omega)\bigg)=\notag\\
\fl&=\frac{\Sigma}{2\pi}\,2\pi\text{i}\bigg(-\sum_i\text{Res} \big[\tilde{R}(-z)f_1(\omega)\big]\bigg\vert_{\omega=\omega_i}+\sum_j\text{Res} \big[\tilde{R}(z)f_2(\omega)\big]\bigg\vert_{\omega=\omega_j} \bigg)
 \end{eqnarray}
where the signs refer to integration contours arranged as 
figures \ref{fig:Isym1} and  \ref{fig:Isym2}, i.e.\ with an anticlockwise orientation.
The functions $f_1$ and $f_2$ are defined as
 \begin{equation}
 \label{eq:f1}
  f_1(\omega)= \frac{1}{z-\text{i}\omega}\,\frac{\frac{\lambda+\text{i}\omega-\sqrt{(\lambda+\text{i}\omega)^2-4\eta}}{2\eta}}{1-\tilde{R}(-z)\bigg(\frac{\lambda+\text{i}\omega-\sqrt{(\lambda+\text{i}\omega)^2-4\eta}}{2\eta}\bigg)}= \frac{1}{z-\text{i}\omega}\,\frac{\tilde{R}(\text{i}\omega)}{1-\tilde{R}(-z)\tilde{R}(\text{i}\omega)}
 \end{equation}
 \begin{equation}
  f_2(\omega)= \frac{1}{z-\text{i}\omega}\,\frac{\frac{\lambda-\text{i}\omega-\sqrt{(\lambda-\text{i}\omega)^2-4\eta}}{2\eta}}{1-\tilde{R}(z)\bigg(\frac{\lambda-\text{i}\omega-\sqrt{(\lambda-\text{i}\omega)^2-4\eta}}{2\eta}\bigg)}=\frac{1}{z-\text{i}\omega}\,\frac{\tilde{R}(-\text{i}\omega)}{\tilde{R}(z)\tilde{R}(-\text{i}\omega)-1}
 \end{equation}
In the last line of \eqref{eq:gen} we have already exploited that, because $\text{lim}_{|\omega|\rightarrow \infty}|\omega f_1(\omega)|=0$ and $\text{lim}_{|\omega|\rightarrow \infty}|\omega f_2(\omega)|=0$, the contributions 
from the semicircles $\mathcal{C}_1$ and $\mathcal{C}_2$ vanish when their radius is sent to infinity.

For further evaluation we first focus on $\eta>0$. As before it is convenient to restrict $z$, here such that it lies outside the left spectral ellipses in figure \ref{fig:symPo}. 
The denominator of $f_1(\omega)$ and $f_2(\omega)$ then has only one relevant zero
 \begin{equation}
  \omega_1=-\text{i}z
 \end{equation}
 This is because with $z$ restricted as above, $|\tilde R(-z)|<1$, for exactly the same reason that $|g_1|<1$ outside the (unshifted) spectral ellipse as discussed in section \ref{sec:ED}. Our choice of contour 1 also guarantees that
 $|\tilde{R}(\text{i}\omega)|<1$ because the integration contour avoids the appropriately rotated spectral ellipse that governs $\tilde{R}(\text{i}\omega)$, as shown in figure \ref{fig:Isym1}. Thus 
 the denominator $1-\tilde{R}(-z)\tilde{R}(\text{i}\omega)$ in \eqref{eq:f1} can never be zero inside our integration contour. An exactly analogous argument applies to the integration over $f_2$. 
 
We now further restrict $z$ to be real and positive, such that $\omega_1$ lies in the lower half plane (see figures \ref{fig:Isym1} and \ref{fig:Isym2}). 
Only the integration contour for $f_1$  then encircles any singularities at all, and we obtain from \eqref{eq:gen}
    \begin{eqnarray}
\tilde{C}(z)&=\Sigma\int_{-\infty}^{+\infty} \frac{d \omega}{2\pi} \big[\tilde{R}(-z)f_1(\omega)+\tilde{R}(z)f_2(\omega)\big]=\notag\\
&=-\frac{\Sigma}{2\pi}\,2\pi\text{i}\,\text{Res}\big[\tilde{R}(-z)f_1(\omega)\big]\bigg\vert_{\omega=\omega_1}=\frac{\Sigma\,\tilde{R}(z)\tilde{R}(-z)}{1-\tilde{R}(z)\tilde{R}(-z)}
 \end{eqnarray}
as claimed in \eqref{eq:fincor} in the main text.
 
\begin{figure}
\begin{center}
\begin{tikzpicture}
\draw [help lines,->] (0,-5) -- (0,5);  
\draw [help lines,->] (-5,0) -- (5,0);   
\node at (5.2,-0.3){$\text{Re}(\omega)$};
\node at (-0.6,4.8) {$\text{Im}(\omega)$};
\node at (-2,-4.5) {$\mathcal{C}_1$};
\node at (0,-0.5) {$\bm{\times}$}; 
\node at (0.5,-0.5){$\omega_1$};
\node at (-4.7,4.7){$I_1$};
\node at (0,2){$\bm{-}$};
\node at (0,3){$\bm{-}$};
\node at (1,2){$\lambda-2\sqrt{\eta}$};
\node at (1,3){$\lambda+2\sqrt{\eta}$};
\draw[very thick,black]
  (0,2) -- (0,3);
\draw[very thick,black,xshift=2pt,
decoration={ markings,  
      mark=at position 0.2 with {\arrow{latex}}, 
      mark=at position 0.6 with {\arrow{latex}},
      mark=at position 0.8 with {\arrow{latex}}, 
      mark=at position 0.98 with {\arrow{latex}}}, 
      postaction={decorate}]
  (-4.5,0) -- (4.5,0);
\draw[thick,black,xshift=2pt,
decoration={ markings,
      mark=at position 0.2 with {\arrow{latex}}, 
      mark=at position 0.4 with {\arrow{latex}},
      mark=at position 0.6 with {\arrow{latex}}, 
      mark=at position 0.8 with {\arrow{latex}}}, 
      postaction={decorate}]
 (4.5,0) arc (0:-180:4.5) ;
\draw[style=dashed] (0,2.5) ellipse (0.7 and 1.5);
\end{tikzpicture}
  \caption{Integration contour in the complex plane for $I_1$ in the case of generic symmetry $\eta>0$.}
\label{fig:Isym1}
\end{center}
\end{figure}
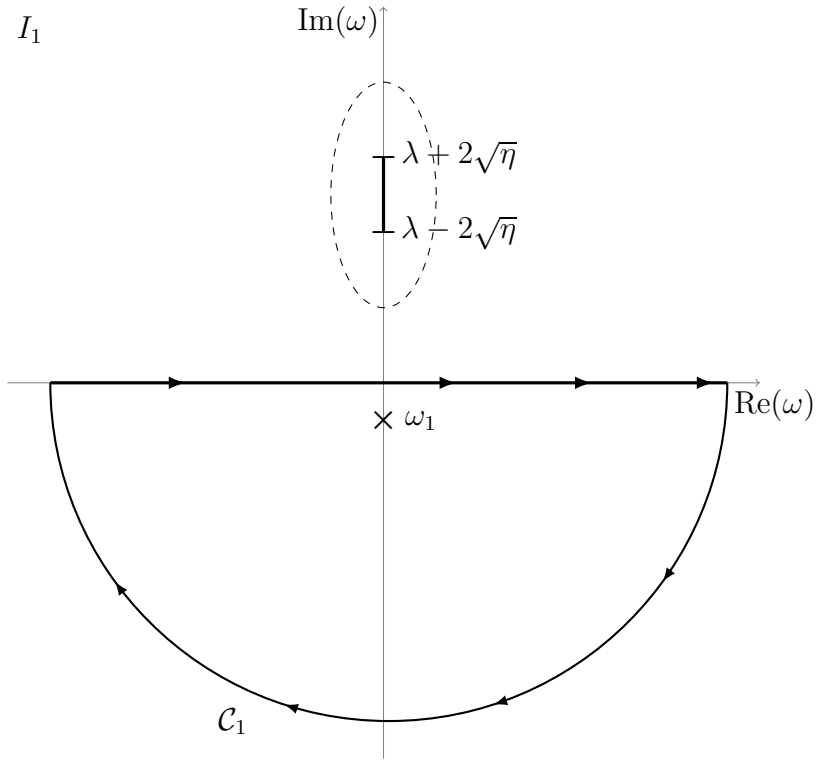

\begin{figure}
\begin{center}
\begin{tikzpicture}
\draw [help lines,->] (0,-5) -- (0,5);  
\draw [help lines,->] (-5,0) -- (5,0);   
\node at (4.8,-0.3){$\text{Re}(\omega)$};
\node at (-0.6,4.8) {$ \text{Im}(\omega)$};
\node at (-2,4.5) {$\mathcal{C}_2$};
\node at (0,-0.5) {$\bm{\times}$}; 
\node at (0.5,-0.5){$\omega_1$};
\node at (-4.7,4.7){$I_2$};
\node at (0,-2){$\bm{-}$};
\node at (0,-3){$\bm{-}$};
\node at (1.2,-2){$-\lambda+2\sqrt{\eta}$};
\node at (1.2,-3){$-\lambda-2\sqrt{\eta}$};
\draw[very thick,black]
  (0,-2) -- (0,-3);
\draw[very thick,black,xshift=2pt,
decoration={ markings,  
      mark=at position 0.2 with {\arrow{latex}}, 
      mark=at position 0.6 with {\arrow{latex}},
      mark=at position 0.8 with {\arrow{latex}}, 
      mark=at position 0.98 with {\arrow{latex}}}, 
      postaction={decorate}]
  (-4.5,0) -- (4.5,0);
\draw[thick,black,xshift=2pt,
decoration={ markings,
      mark=at position 0.2 with {\arrow{latex}}, 
      mark=at position 0.4 with {\arrow{latex}},
      mark=at position 0.6 with {\arrow{latex}}, 
      mark=at position 0.8 with {\arrow{latex}}}, 
      postaction={decorate}]
 (4.5,0) arc (0:180:4.5) ;
\draw[style=dashed] (0,-2.5) ellipse (0.7 and 1.5);
\end{tikzpicture}
  \caption{Integration contour in the complex plane for $I_2$ in the case of generic symmetry $\eta>0$.}
\label{fig:Isym2}
\end{center}
\end{figure}

For $\eta<0$ an analogous calculation of the correlation function integral can be performed. Some changes in the relevant regions of the complex plane occur, 
namely the ellipses bounding the support of the spectrum are rotated (compare figures \ref{fig:symPo} and \ref{fig:symPoN}), but the method is the same for $\eta>0$, and so is the result.

\cleardoublepage

\section{Complete TAP equations}
\label{sec:complete_TAP}
We lift the restriction ${\partial \phi_i(\bm{x}(t))}/{\partial x_i(t)}=0$ and the one regarding the additivity of variables in the drift $\phi_i(\bm{x}(t))$.
Then the dynamical equation up to $\alpha^2$ order can be written, from \eqref{eq:zerofields} and \eqref{eq:eff_dyn}, in the form
\begin{eqnarray}
\label{eq:ovdyn}
\fl\frac{d x_i(t)}{dt}=& -\lambda_i x_i(t)+ \alpha\langle\phi_i(t)\rangle
+\alpha \langle\partial_i \phi_i(t)\rangle\delta x_i(t)
\notag\\
&{}+\alpha^2\bigg[\int_{0}^t dt'\sum_j\, R_j(t,t')\langle \partial_{j}\phi_i(t)\partial_{i}\phi_j(t')\rangle\delta x_i(t')\notag\\
&{}-\int_{0}^t dt'R_i(t,t')\langle\partial_i\phi_i(t)\rangle\langle \partial_i \phi_i(t')\rangle\delta x_i(t')\notag\\
&{}+\int_{0}^t dt'\,\sum_j R_j(t,t')(\langle\partial_j\phi_i(t)\delta \phi_j(t')\rangle-\langle \partial^2_j \phi_i(t) \rangle C_j(t,t')\langle \partial_j\phi_j(t')\rangle)\notag\\
&{}+\int_{0}^t dt'\,\sum_j \big(\langle\partial_i\partial_j\phi_i(t)\delta \phi_j(t')\rangle-\delta C_j(t,t')\langle \partial_j \phi_j(t')\rangle \langle\partial_j^2\partial_i \phi_i(t)\rangle\big)\notag\\
&{}\qquad\qquad \qquad  R_j(t,t')\delta x_i(t)\bigg]+\xi_i(t)+\chi_i(t)
\end{eqnarray}
where the effective noise has correlator
\begin{equation}
\label{eq:beffself}
\langle \chi_i(t)\chi_i(t')\rangle= \frac{\alpha^2}{2}\big(\langle \delta\phi_i(t)\delta \phi_i(t')\rangle-\langle\partial_i\phi_i(t)\rangle\langle \partial_i \phi_i(t')\rangle\delta C_i(t,t')\big)
\end{equation}
For the sake of brevity we have dropped all $\bm{x}$-dependencies above, writing e.g.\ $\partial_i \phi_i(t)={\partial \phi_i(\bm{x}(t))}/{\partial x_i(t)}$ and $\phi_i(t)=\phi_i(\bm{x}(t))$. 

\newcommand{\dtp}{\delta\tilde{\phi}}

Compared to \eqref{eq:dx_dt_TAP} in the main text, there are a number of additional terms. The last term in the first line is the linearization of the self-interaction already familiar 
from our generic first order result \eqref{eq:dxi_dt_first_order}. This systematic effect of the self-interaction is correspondingly removed from the effective noise $\chi_i(t)$, whose correlator \eqref{eq:beffself} is easily shown to be the correlation function of $\dtp_i\equiv \delta\phi_i-\langle \partial_i \phi_i\rangle \delta x_i$. This is the genuinely interacting part of the drift, i.e.\ the one that is 
not captured in the first line of \eqref{eq:ovdyn}. The third line of \eqref{eq:ovdyn} similarly subtracts off the self-interaction term from the main memory term in the second line.

The fourth line of \eqref{eq:ovdyn} is a contribution that is independent of the specific history of $x_i$; instead it involves a time integral of {\em averages} over fluctuation statistics in the past. It can again be written in terms of $\dtp_i$, with the coefficients in brackets after 
$R_j(t,t')$ equal to $\langle \partial_j \phi_i(t)\dtp_j(t')\rangle$. The coefficient in front of $\delta x_i(t)$ in the fifth and sixth line of \eqref{eq:ovdyn} has an analogous form, as $\langle \partial_i \partial_j \phi_i(t)\dtp_j(t')\rangle$.
The overall contribution from the fourth, fifth and sixth line of \eqref{eq:ovdyn} can be cast as
\begin{equation}
\alpha^2 \int_{0}^t dt' \sum_j R_j(t,t')\big(\langle \partial_j \phi_i(t)\dtp_j(t')\rangle
+\delta x_i(t)\langle \partial_i \partial_j \phi_i(t)\dtp_j(t')\rangle\big)
\end{equation}
This has a fairly straightforward interpretation: a fluctuation in the drift of variable $j$ ($\dtp_j(t')$) that changes $x_j(t')$ is propagated forward to time $t$ by $R_j(t,t')$ and then affects the drift $\tilde\phi_i(t)$ including the linearized dependence on $x_i$. 

It is interesting to note that all of the additional terms disappear if there are no self-interactions ($\partial_i \phi_i=0$), except for the first term in the fourth line of \eqref{eq:ovdyn}. 
The latter vanishes if one makes in addition the assumption that interactions are additive in the variables, as then $\partial_j \phi_i$ depends only on $x_j$ and so is independent 
of $\dtp_j$ if there are no self-interactions. In the generic case of non-additive interactions, the first term in the fourth line remains. In particular, it gives a correction to the time evolution of the means
\begin{equation}
\label{eq:ovdynm}
\frac{d \mu_i(t)}{dt}= -\lambda_i \mu_i(t)+ \alpha\langle\phi_i(t)\rangle+\alpha^2\int_{0}^t dt'\,\sum_j R_j(t,t')\langle \partial_j \phi_i(t)\dtp_j(t')\rangle
\end{equation}
In an exact theory, only the first two terms are present, so that the last one has to be interpreted as correcting for the fact that the Plefka expansion produces an 
approximating distribution where all variables are decoupled. For the case of additive interactions, no such correction appears because $\langle \phi_i\rangle$ is then a 
sum of averages over single variables.

For the special case of the $p$-spin spherical model, we have verified that the above equations reproduce those derived by other means by Biroli \cite{biroli}. The correction to the mean dynamics vanishes for $p=2$ as expected, as the interactions are then additive, but is nonzero for $p>2$ where the drift involves products of variables.
   
\cleardoublepage
\bibliography{./Plefkabib}
\nocite{*}

\end{document}